\documentclass[]{article}
\usepackage[british]{babel}
\usepackage[utf8]{inputenc}
\usepackage{graphicx}
\usepackage{color}
\usepackage{algorithm}
\usepackage{algpseudocode}
\usepackage{amsmath}
\usepackage{comment}
\usepackage{xcolor}
\usepackage{hyperref}
\usepackage{authblk}
\usepackage{subcaption}
\usepackage{geometry}
\usepackage[export]{adjustbox}
\usepackage{caption}
\usepackage[sorting=none]{biblatex}
\usepackage{multicol}
\usepackage{float}
\usepackage{dblfloatfix}
\usepackage{xr}

\title{Car dependency in access to opportunities in cities}
\title{Evaluating Car Dependency and Access to Opportunities in Urban Environments}
\title{How Urban Accessibility Shapes Car Dependency}
\title{Car Dependency in Urban Accessibility}

\author[1,2]{Bruno Campanelli\thanks{Corresponding author: campanelli.bruno@gmail.com}}
\author[1,2,3]{Francesco Marzolla}
\author[1,2]{Matteo Bruno}
\author[1,2,4,5]{\\Hygor P.M. Melo}
\author[1,2,3,6]{Vittorio Loreto}

\affil[1]{Sony Computer Science Laboratories - Rome, Joint Initiative CREF-SONY, Centro Ricerche Enrico Fermi, Via Panisperna 89/A, 00184, Rome, Italy}
\affil[2]{Centro Ricerche Enrico Fermi (CREF), Via Panisperna 89/A, 00184, Rome, Italy}
\affil[3]{Sapienza Univ. of Rome, Physics Dept, Piazzale A. Moro, 5, 00185, Rome, Italy}
\affil[4]{Postgraduate Program in Applied Informatics, University of Fortaleza, 60811-905, Fortaleza, CE, Brazil}
\affil[5]{Núcleo de Ciência de Dados e Inteligência Artificial (NCDIA), Univ. of Fortaleza, 
60811-905, Fortaleza, CE, Brazil}
\affil[6]{Complexity Science Hub, Metternichgasse 8, A 1030, Vienna, Austria}

\date{}

\begin{document}

\maketitle

\begin{abstract}

To achieve net-zero emissions, cities must transition away from reliance on private vehicles. However, car-centric urban growth has transformed the automobile from a convenience tool into a necessity for accessing essential services, creating significant ``car dependency". This study introduces a novel Car Dependency Index (CDI) that quantifies the accessibility gap between private and public transport across 18 cities in Europe and North America. Utilising high-resolution geospatial data and numerical simulations, we reveal pronounced spatial inequalities, showing that car dependency remains a primary driver of car ownership even when accounting for income. A ``what-if" simulation of the planned metro expansion in Rome predicts a reduction of approximately 60,000 commuting vehicles, yet highlights that isolated interventions have localised impacts. We conclude that systemic, network-level transit expansions are essential to dismantle car-based systems and foster equitable, sustainable urban mobility. Our framework provides policymakers with an objective, scalable tool to identify viable areas for car-free zones and target infrastructure investments effectively.

\end{abstract}

\begin{multicols}{2}
\section*{Introduction}

The movement toward sustainable urban living is transforming cities and is centred on a debate on the role of cars in daily life. To advance sustainable transportation, urban planners, policy makers, and activists implement measures such as congestion charges~\cite{borjesson2012stockholm,prud2005london}, car-free days~\cite{glazener2022impacts}, cycling infrastructure~\cite{basilone2025}, and pedestrian zones~\cite{rhoadsInclusive15minuteCity2023}. Additional strategies include limiting parking~\cite{albalate2020impact}, improving public transport~\cite{nieuwenhuijsen2016car}, raising residential parking costs~\cite{ostermeijer2019residential}, and establishing car-free areas~\cite{rueda2019superblocks}.

Reducing car usage is central to these measures, as it is crucial to obtain net zero carbon emissions~\cite{NetZero2050, IEA2023} while benefiting air quality~\cite{wallington2022vehicle}. They also address the often-overlooked burden of deaths and injuries from road crashes~\cite{peden2004world,folco2023data}. Other benefits include less noise, less congestion, reduced urban heat effects~\cite{gossling2020cities}, and more public space that would otherwise be needed to move and park private cars. These changes help build more liveable urban environments~\cite{verkade2024movement, brommelstroetIdentifyingNurturingEmpowering2022}. However, the role of the car is a nuanced issue.

On the other hand, cars have long helped create more equal opportunities in cities~\cite {blumenberg2014driving}. They give many people reliable ways to get around. In this view, cars are not a dependency but a tool for fair access, even if they do not always save time~\cite{metz2008myth}. However, urban growth has become increasingly dependent on cars~\cite{newman1989cities}. This has shifted cars from a means to reach more options to a dependency. Without them, many people would not have access to even basic services~\cite{lucas2012transport,urry2004system}. Lately, this situation has been reconsidered to find new, fair ways of access. The push comes from the need for sustainable solutions and from problems caused by more people moving to cities, which worsen traffic~\cite {chang2017there,loufHowCongestionShapes2014}. This traffic slows things down and lowers quality of life~\cite{han2022effect}. Even though reducing car use has clear environmental and social benefits, it is still hard to move beyond car-based systems. This is because of the strong political and economic factors supporting them~\cite{mattioli2020political}, which leads to resistance to restricting or removing cars from cities.

Understanding the spatial patterns of access that generate car dependency is essential for designing effective and equitable transport policies. The opportunities available to individuals depend mainly on proximity and ease of transportation~\cite {bertolini1999spatial,bruno2026dimensions}, and studies show that having essential services within walking or cycling distance reduces emissions~\cite{bruno2024universal,marzolla2024compact,marzolla2026proximity}. However, proximity alone is insufficient in large cities, where not all destinations—such as workplaces or higher-quality services—can be nearby~\cite{abbiasov202415,hill2024beyond}. As a result, the efficiency and convenience of transportation modes remain crucial for accessing opportunities. Urban areas vary in their reliance on private vehicles: income, land-use patterns, and the availability of alternative modes shape this reliance. Recognising these spatial disparities enables policymakers to target interventions more precisely, to ensure efforts to reduce car use do not reinforce social or spatial segregation but instead promote inclusive and sustainable access for all residents. 

Recent studies therefore analyse inequalities in accessibility between private and public transport. Mattioli~\cite{mattioli2017forced} highlights how car dependency intersects with socioeconomic distress, while Wiersma et al.~\cite{wiersma2021spatial} examine the spatial determinants of dependency, showing how density, land-use mix, and accessibility influence modal choice. Wu et al.~\cite{wu2021urban} adopt a similar approach by comparing job accessibility by car and public transport across metropolitan areas worldwide. Most recently, Kiberd and Straňák~\cite{kiberd2024CarDependency} propose a car dependency index for English cities that integrates indicators of potential accessibility, commuting behaviour, and residential density.

Building on these insights, we quantify the spatial distribution of car dependency in cities using a comprehensive accessibility framework. This framework compares opportunities reachable by car and by public transport, using open data. We develop a fine-grained index based solely on potential access to essential services and leisure activities, capturing each area’s capacity to support car-free living. We frame cities as \textit{cities of opportunities}~\cite{hill2024beyond,bruno2026dimensions} to show how transport provision and residential location interact, creating systematic inequalities between car owners and public transport users. We apply this framework to 18 cities across Europe and North America to measure accessibility gaps within and between cities. We validate the index against observed commuting behaviour and link car-free potential to income and car ownership in Vienna. Finally, we simulate planned metro infrastructure in Rome to estimate its impact on car dependency and potential reductions in car use, illustrating how public transport investments can lower reliance on private vehicles.

Together, by identifying areas where car removal is feasible, our contributions provide an objective, scalable measure of car dependency that enables more equitable, evidence-based urban mobility policy.

\section*{Methods}

\begin{figure}[H] 
\centering
    \includegraphics[width=\columnwidth]{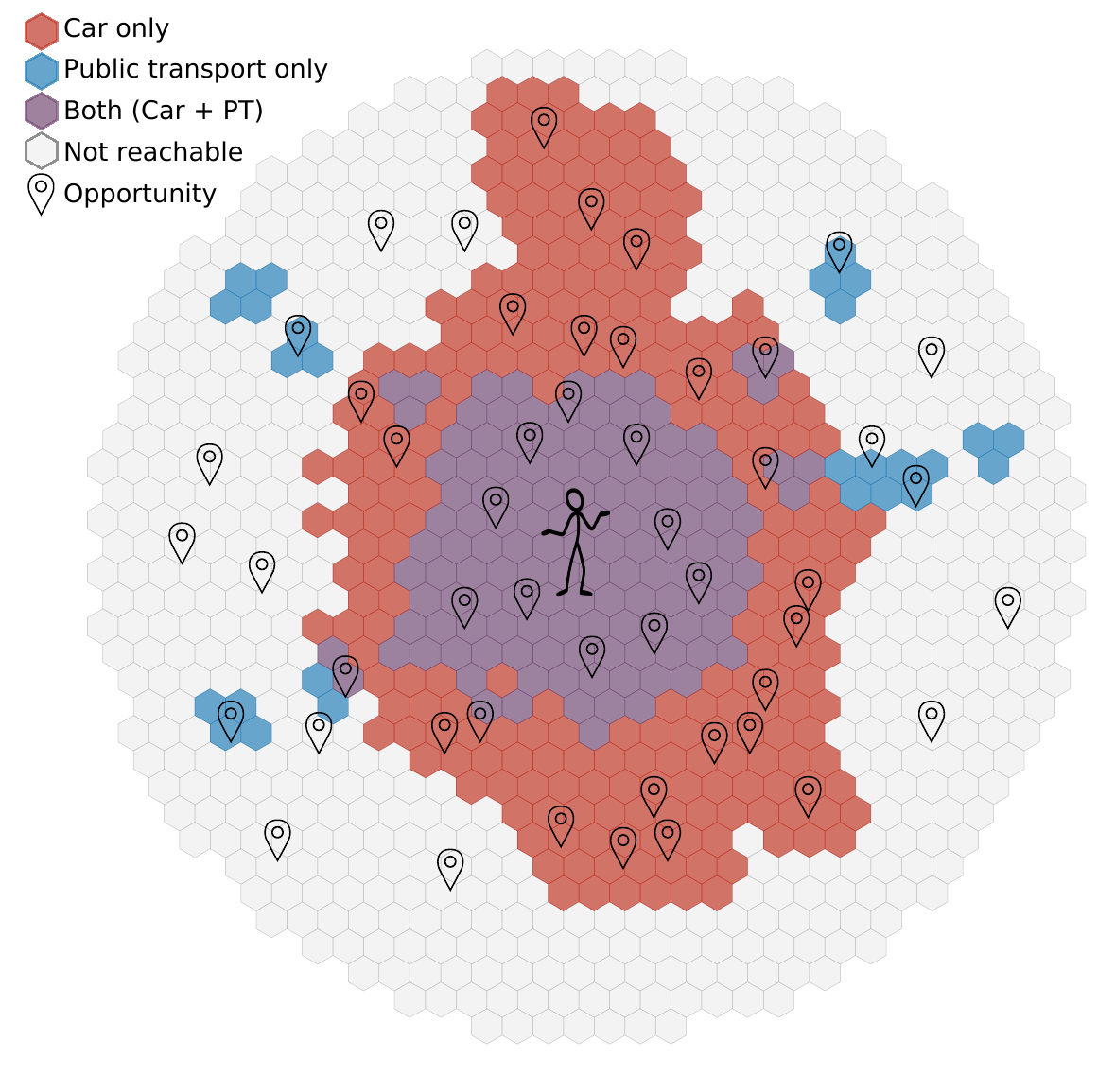}
    \caption{\textbf{Schematic representation of POIs accessibility by different transport modes.} The agent at the centre of the figure can access points of interest (POIs), indicated by pins, using either public transport or a private car. Areas of the city, represented on a hexagonal grid, are shown in red if reachable by car within a normal-length journey, in blue if reachable by public transport within a normal-length journey, and in purple if reachable by both modes.}
    \label{fig:omino}
\end{figure}

\subsubsection*{Static data: city borders and quantities}
For each city, we define the study area based on its official administrative boundaries, downloaded from the municipal websites at the time of data analysis. This choice reflects data availability constraints, as obtaining consistent public transport information for peripheral areas is more complex due to its multiple sources and could bias our car dependency estimates. In our framework, each city consists of all the resolution-9 hexagons from the H3 geospatial indexing system~\cite{h3geo} whose centroids fall within the city limits. This resolution corresponds to a hexagon side length of approximately 200 m.

The population residing inside each hexagon is derived from the gridded population data provided by WorldPop~\cite{WORLDPOP}.
The latter are defined on a square grid with a side of 100 m. We assume the population within each square is uniformly distributed. The population of a hexagon $h$ is calculated as the sum of the population counts of the intersections between $h$ and the squares. Hexagons with a population count of zero are excluded from the analyses and shown in grey in all plots. All aggregate results shown are averaged using population as a weight.

The Points Of Interest (POIs) inside each city are obtained from OpenStreetMap~\cite{OSM} data downloaded from~\cite{geofabrik}. Data objects can be points or polygons; in that case, we use the centroid as the POI's location. All entries have attached tags, consisting of key:value pairs, such as amenity:restaurant. We manually built a list of tags identifying POIs and discarded data entries that did not contain any of those tags. Each POI is then assigned to the hexagon that contains it.

\subsubsection*{Travel times computation}

Travel times by public transport and by private car are computed for all origin-destination pairs in a city. For the sake of tractability, all paths begin and end at the centroid of a hexagon, and we assume that if a destination centroid is reachable within a time $t$, then so are all POIs within the destination hexagon. For both public transport and cars, we assume the fastest path between the origin and destination is chosen. In the case of public transport, we first compute the travel times by foot between all hexagon pairs with a line-of-flight distance of less than 1.25 km (corresponding to travelling in a straight line without obstacles at a speed of 5 km/h for 15 minutes) using the Open Source Routing Machine (OSRM)~\cite{OSRM} with OpenStreetMap data. Together with the public transport schedule, the foot times are used as input for the Connection Scan Algorithm~\cite{CSA}. This produces travel times for the fastest public transport routes that allow walking for up to 15 minutes between stops (e.g., getting out of a metro station to take a bus). Car times are also computed through OSRM, and corrected as:
\begin{equation}
    t_{\text{Car}}(o, d) = (1 + \Delta_{\text{City}}^\text{Traffic})\,t_{\text{Car}}^\text{OSRM}(o, d) + 15\min
    \label{eq:cart}
\end{equation}
where $\Delta_{city}^\text{Traffic}$ is the city's average time lost to traffic in 2023 compared to driving in free-flow conditions, computed by TomTom in their 2023 Traffic Index Report~\cite{tomtom}, $t_{\text{Car}}^\text{OSRM}(o, d)$ is the “raw” travel time for the fastest path between $o$ and $d$ by car as computed by OSRM, and we added 15 minutes to account for walking to the car before the trip, finding a parking spot, and walking from the parking spot to the actual destination. This buffer is arbitrary and depends on many factors, such as the city, urban areas, and time of day, and is intended to make the simulation more reflective of real-life conditions. While the extra time required can vary, adjusting this parameter would merely shift our overall metrics without affecting the relative territorial differences.

\subsubsection*{Opportunity metrics}

We now define $H_{h,m}(t_0, t)$ as the set of hexagons whose centres can be reached using
mode of transportation $m \in \{\text{PT}, \text{Car}\}$, starting from the centre of hexagon $h$ at time $t_0$ and travelling for $t$, and $P_{h,m}(t_0, t)$ as the number of POIs contained in $H_{h,m}(t_0, t)$ (note that time here has a resolution of one second). In Fig.~\ref{fig:omino} we provide a schematic figure of our framework. In terms of this, we define the \textit{opportunity score} of a hexagon as:

\begin{equation}
    O_{h,m} = \left\langle\int_{t=t_0}^\infty dt \, f(2t) \, P_{h,m}(t_0, t)\right\rangle_{t_0} \;,
\end{equation}

where the average is taken over $t_0$ from 8 AM to 10 PM in steps of 1 hour. Night hours are excluded for simplicity, as they would require separate treatment due to different PT scheduling and POI usage patterns. This quantity can be seen as one of the possible implementations of the \textit{primal access} metric~\cite{hansen1959accessibility,levinsonGeneralTheoryAccess2020}. Here $f(t)$ is a utility function such that $f(t)\rightarrow 0$ for $t \rightarrow \infty$ and
$\int f(t) \, dt = 1$, representing the idea that POIs farther from the hexagon centre weigh less on the opportunities of residents as they need more time to be reached, and is calculated in $2t$ to account for the return trip. In the results section, we used:

\begin{equation}
    f(t) = \frac{1}{\tau} e^{-t/\tau}
\end{equation}

with $\tau$ equal to 60 minutes. We verified (see supplementary material) that Spearman's rank correlation coefficients between results obtained with different reasonable choices for $f$ are close to 1.

If one thinks about $f(t)$ as the probability density of doing a journey of duration $t$, then $O_{h,m}$ can be interpreted as the average number of POIs (”opportunities”) that a resident of $h$ can encounter during a typical day.

\subsubsection*{Car Dependency Index metric}

We now define the Car Dependency Index $\text{CDI}_h$ as:
\begin{equation}
    \text{CDI}_h = \frac{O_{h,\text{Car}}-O_{h,\text{PT}}}{O_{h,\text{Car}}+O_{h,\text{PT}}} \,.
\end{equation}
Positive (negative) values of $\text{CDI}_h$ indicate that opportunities are more easily accessible by car (public transport). While the values and sign of $\text{CDI}_h$ may vary with the chosen parameters and metrics, the relative inequalities within and between cities remain consistent across comparable specifications. Averaging over all hexagons using their population counts as weights, we get the city's opportunity scores $O_m$ and the global car dependency index $\text{CDI}$.

\subsubsection*{Impact of car dependency on car usage}

In studying the impact of car dependency on car usage, we use data on modal share for commuting collected from various surveys~\cite{PRIETOCURIEL2024ABC}. From the city populations reported in the dataset, we deduce that the data for the cities under study refer to the portions of those cities enclosed by administrative boundaries. We evaluate the correlation between the population-weighted median of the CDI and the percentage of commuting trips made by private vehicles in each city.  

\subsubsection*{Impact of car dependency on car ownership}

We then use Vienna as a case study to explore how car dependency and car ownership interact.
To estimate the fraction of Vienna's citizens with access to a car, we assume that if a household member owns a car, all other household members have access to it. In doing so, we also consider that underage people could access it, implying that the adults who take care of them would drive them to the POIs where they can access opportunities. The household-size distribution in Austria~\cite{household_size} is considered valid in Vienna for simplicity as well. We know from the municipality of Vienna~\cite{Vienna} the number of registered vehicles $V$ per district. We also know the resident population $P$ of each district~\cite{Vienna_censo}. We therefore can compute the motorisation $M$ of each district as:
\begin{equation}
    M = \frac{V}{P} \, .
\end{equation}
The probability of not having a car, for a resident extracted randomly from a district with motorisation $M$ is: 
\begin{equation}
    P(\text{car-less person})(M) = (1-M) \, .
\end{equation}
Therefore, assuming that everyone buys cars independently from the other people in their household, the probability that no one in a household of $n$ people owns a car is:
\begin{equation}
    P(\text{car-less hous.})(M,n) = \left( 1-M \right)^n \, .
\end{equation}
Then the probability that at least one in the household has a car, and therefore it's shared with all the members of the household, is:
\begin{equation}
    P(\text{car hous.})(M,n) = 1- P(\text{car-less hous.})(M,n)\, .
\end{equation}
Then, finally, the fraction $F$ of people having access to a car in a district with motorisation $M$ is:
\begin{equation}
\begin{split}
        F (M) &= P(\text{car person}) (M) = \\ 
    &= \frac{\sum_n P(\text{car hous.})(M, n) \cdot n \cdot f(n)}{\sum_n n \cdot f(n)} \, ,
\end{split}
\end{equation}
where $f(n)$ is the frequency of households with $n$ components in Austria~\cite{household_size}.

The entire case study in Vienna is conducted at the district level. To compute the car dependency index CDI of a district, we start from the hexagonal grid previously discussed and assign to each district the average value of CDI of hexagons whose centroids fall within the district's boundaries, weighting each hexagon by its population.

\begin{figure}[H] 
    \includegraphics[width=0.9\columnwidth]{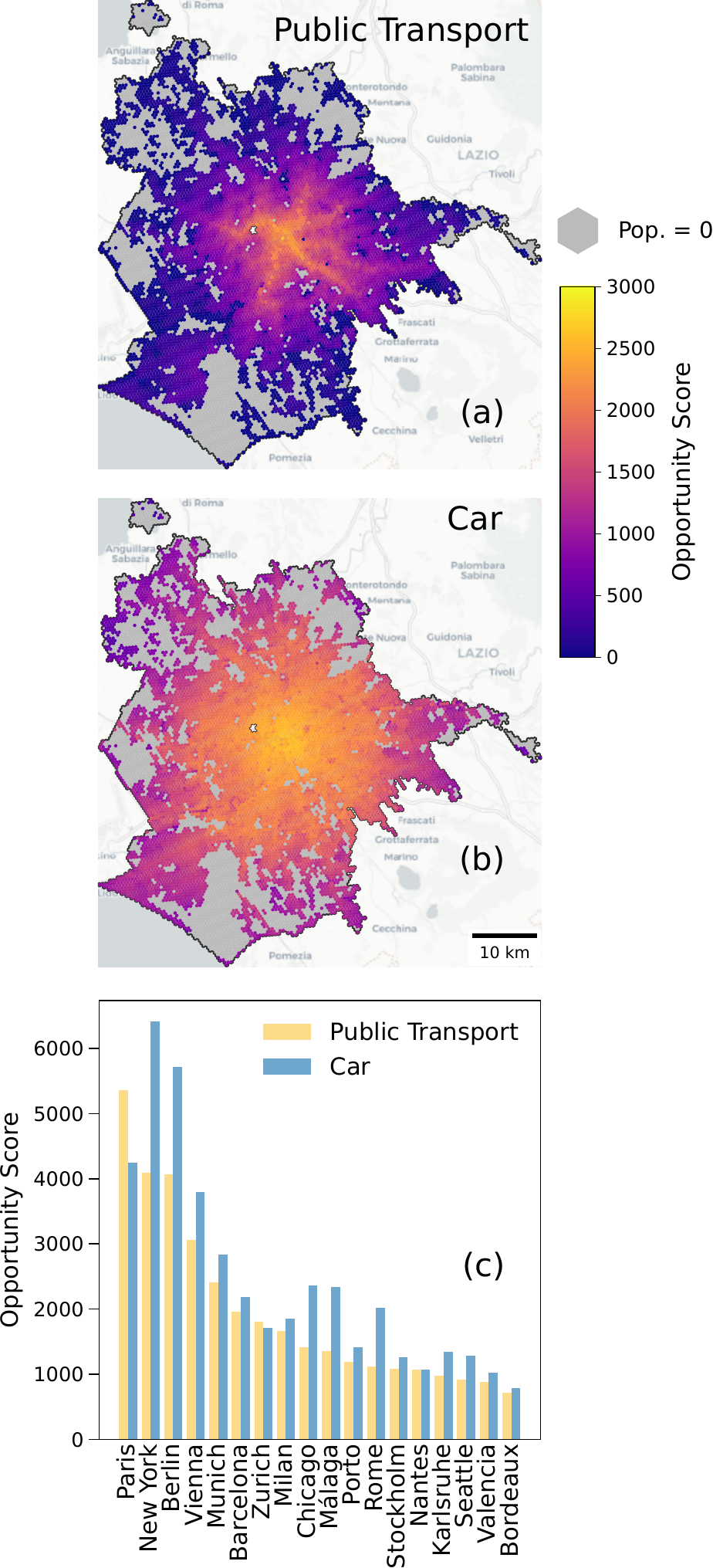}
 \caption{\textbf{Opportunity scores by public transport and car.}  Panels (a) and (b) present Opportunity Scores in Rome by public transport and by car, respectively. Panel (c) shows the population-averaged opportunity scores of various cities.}
     \label{fig:opp}
\end{figure}

\section*{Results}
\begin{figure}[H] 
\centering
    \includegraphics[width=\columnwidth]{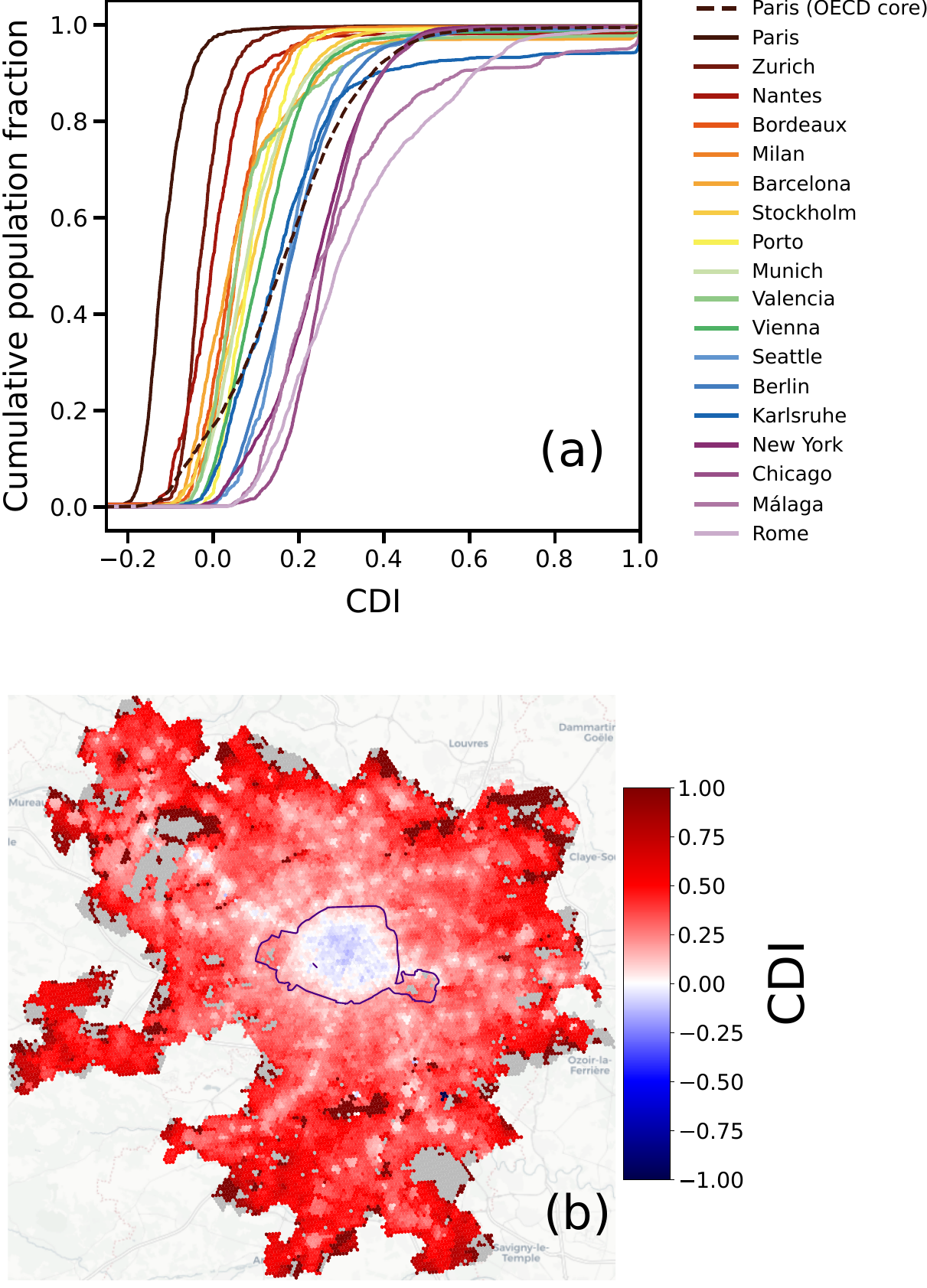}
    \caption{\textbf{Car Dependency Index distribution among city residents.} In Panel (a), we show the cumulative distribution of car dependency scores for the resident population of different cities. In Panel (b), as an example, we show the spatial distribution of the Car Dependency Index values in the Paris metropolitan area, highlighting the smaller administrative boundary.}
    \label{fig:city_statistics}
\end{figure}

Fig.~\ref{fig:opp} illustrates the difference in accessibility between car and public transport. The maps of Rome display, in colour, the opportunity scores by public transport (Panel a) and car (Panel b). Rome is one of the cities in the analysis where this contrast is most pronounced, and the map clearly shows that the two numbers are comparable only in areas near the metro lines, whose network is clearly visible in panel (a). Opportunities reachable by car remain higher in the centre, but the core-periphery divide is lower. Among all the cities in our sample, only Paris and Zurich have a higher opportunity score for public transport than for cars, as shown in the histogram in panel (c). These aggregated scores should be taken as general indications, and it is worth noting that both have relatively small administrative boundaries.
\begin{figure*}[htbp] 
\centering
    \includegraphics[width=\textwidth]{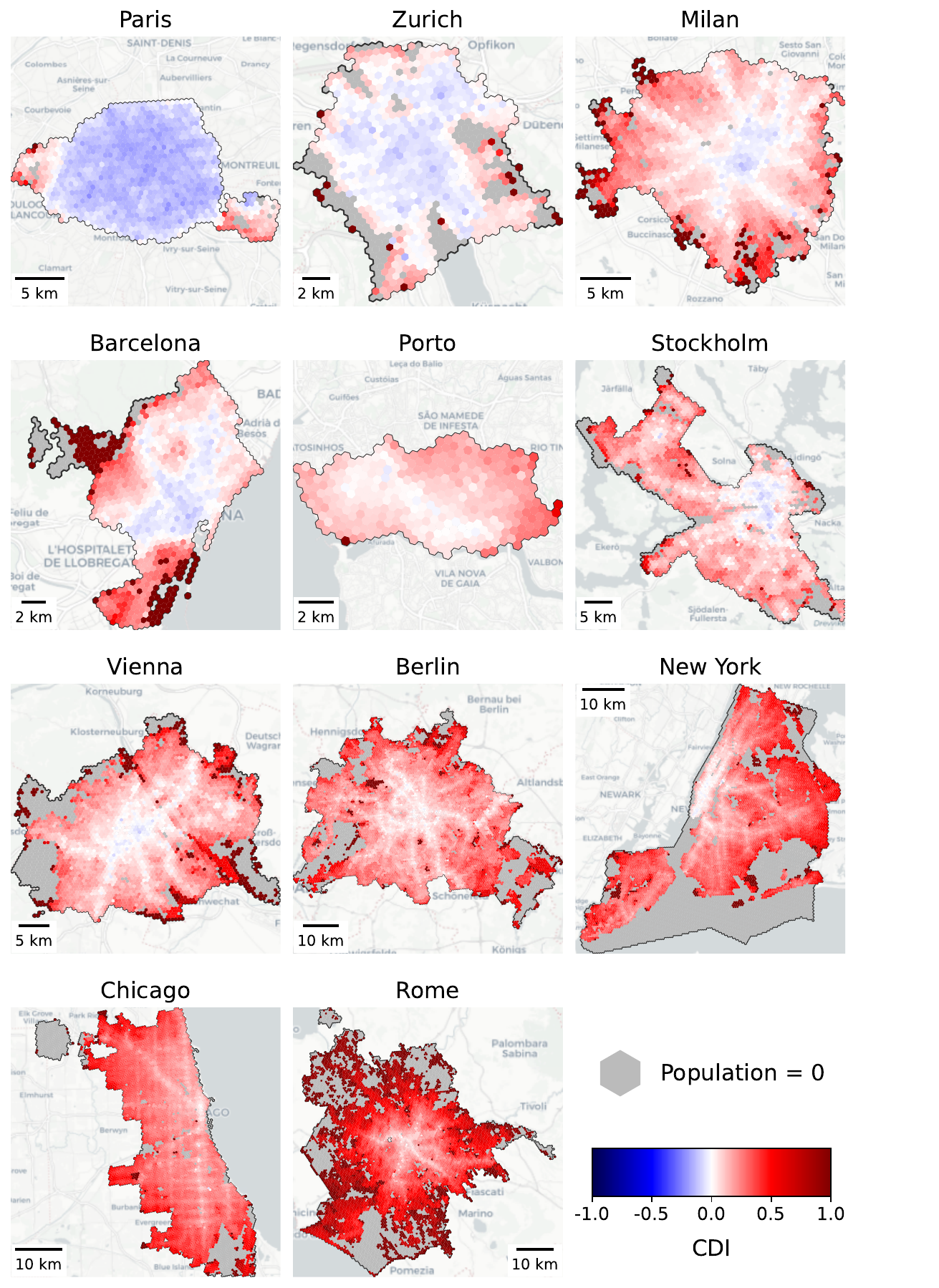}
        \caption{\textbf{Maps of Car Dependency Index.} In the blue areas, public transport is estimated to be more efficient than cars, while the opposite is true in the red areas. Peripheries are generally more car-dependent, but the presence of urban train stations and metro stops has a positive impact—evident in the white spots where a car is not strictly necessary.}
        \label{fig:CDI_cities}
\end{figure*}
Fig.~\ref{fig:CDI_cities} presents maps for a selected subset of the cities under study, illustrating the spatial distribution of the Car Dependency Index (CDI). Maps for all 18 cities are available in the supplementary materials or online\footnote{https://mat701.github.io/CDI/}. Negative CDI values are represented by shades of blue, positive values by shades of red, while hexagons with no resident population are displayed in grey. Analogous plots for all the cities under study are available in the supplementary material. In general, central areas have lower car dependency than peripheral areas, but urban rail infrastructure, such as metro lines, extends low-car dependency into the city's outskirts. Table~\ref{table:CDI_values} lists the average CDIs of all cities under study.
\begin{table}[H]
\centering
\setlength{\tabcolsep}{6pt} 
\renewcommand{\arraystretch}{1.2} 
\caption{\textbf{Average Car Dependency Indices of cities}}
\begin{tabular}{ccc}
\textbf{Rank} & \textbf{City} & \textbf{CDI}  \\
\hline
1 & Paris (Municipality) & -0.111  \\
2 & Zurich & -0.020 \\
3 & Nantes & 0.013  \\
4 & Bordeaux & 0.053  \\
5 & Milan & 0.063 \\
6 & Barcelona & 0.087  \\
7 & Porto & 0.091  \\
8 & Stockholm & 0.094  \\
9 & Munich & 0.102 \\
10 & Valencia & 0.102 \\
11 & Vienna & 0.129 \\
12 & Paris (OECD City) & 0.166 \\
13 & Seattle & 0.188 \\
14 & Berlin & 0.195 \\
15 & Karlsruhe & 0.202 \\
16 & New York & 0.241 \\
17 & Chicago & 0.270 \\
18 & Málaga & 0.310 \\
19 & Rome & 0.335  \\
\end{tabular}
\label{table:CDI_values}
\end{table}
The cumulative CDI distribution by resident population for each city is shown in Fig.~\ref{fig:city_statistics}. For each city, for a given CDI value, the curves show the number of residents living in areas with a car dependency index less than or equal to that value. Therefore, the steeper the curve, the more homogeneous the distribution of the CDI. Cities with lower average car dependency also tend to exhibit lower levels of inequality.

Panel (b) of the same figure reports the CDI computed for the Paris urban core, defined according to OECD~\cite{FUA_OECD}. This area is considerably larger than the municipality, yet smaller than the entire Functional Urban Area. The city centre, located within the municipal boundaries, exhibits substantially lower scores than the surrounding peripheral areas. Consequently, city-wide averages and rankings depend on the spatial delineation adopted, as further illustrated in Table~\ref{table:CDI_values}, where the smaller and larger borders of the city result in substantially different city-wide averages. Border effects instead introduce only negligible variations in the results (see Supplemental Material).
\subsubsection*{Impact of car dependency on car usage}

\begin{figure}[H] 
\centering
\includegraphics[width=.9\columnwidth]{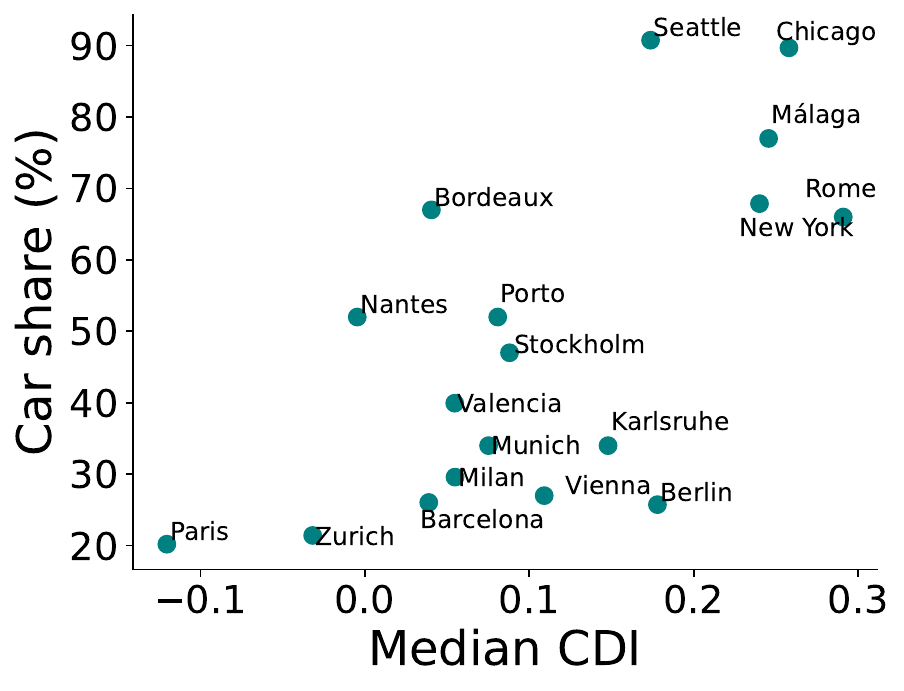}
    \caption{\textbf{Share of car use in commuting vs median car dependency index of the cities considered.}}
    \label{fig:car_use_vs_dependency}
\end{figure}

The car dependency index can shed light on car use in cities. Fig.~(\ref{fig:car_use_vs_dependency}) illustrates the relationship between the median CDI and the percentage of commuting trips made by private vehicles in the cities under consideration. Unsurprisingly, the two variables are positively correlated, with a correlation coefficient of $r = 0.66$ ($p<0.01$). The greater the need (i.e., a larger CDI), the more people opt to own a private car. 

\subsubsection*{Impact of car dependency on car ownership}

\begin{figure*}[!htbp] 
    \centering
    \includegraphics[width=\textwidth]{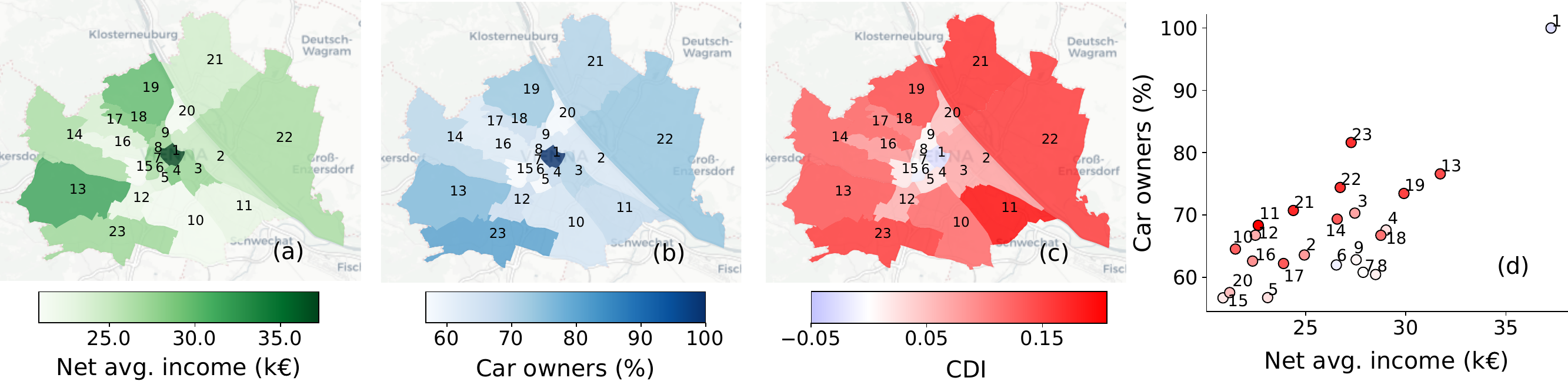}
    \caption{\textbf{Relation among income (a), share of residents potentially using a car (b) and car dependency (c), in Vienna, represented in a scatter plot (d).} Income and car dependency of your area both contribute to your likelihood of being a car owner. Here, car ownership is also intended to be shared: if a household owns a car, every household member is considered a car owner.}
    \label{fig:Vienna_districts}
\end{figure*}

In Fig.~\ref{fig:Vienna_districts}, panel (a) shows the average net income of district residents, panel (b) displays the estimated share of citizens with access to a car across Vienna’s districts, and panel (c) reports the average CDI.

In the historical city centre (district no.~1), the share of residents with access to a car reaches 100\%, which is likely an overestimate. This occurs because the number of cars registered in the district is nearly equal to its population, as many companies have their headquarters—and thus their company cars—there~\cite{Vienna}.

Panel (d) of Fig.~\ref{fig:Vienna_districts} summarises the relationship between car dependency, car ownership, and wealth in Vienna. Each dot represents a district (labelled by number), with the x-axis showing average net income and the y-axis the share of residents with access to a car. The colour scale indicates the district’s average CDI. Data for District no.~1 should be interpreted cautiously due to the bias introduced by company car registrations. The remaining districts reveal a clear trend: wealthier districts tend to have higher car ownership, but for a given income level, those with a greater CDI value show a greater share of car owners. This pattern underscores the combined influence of economic means and neighbourhood car dependency on car ownership decisions.

\subsubsection*{Impact on car dependency of a new public transit infrastructure}

\begin{figure*}[!htbp] 
    \centering
    \includegraphics[width=\textwidth]{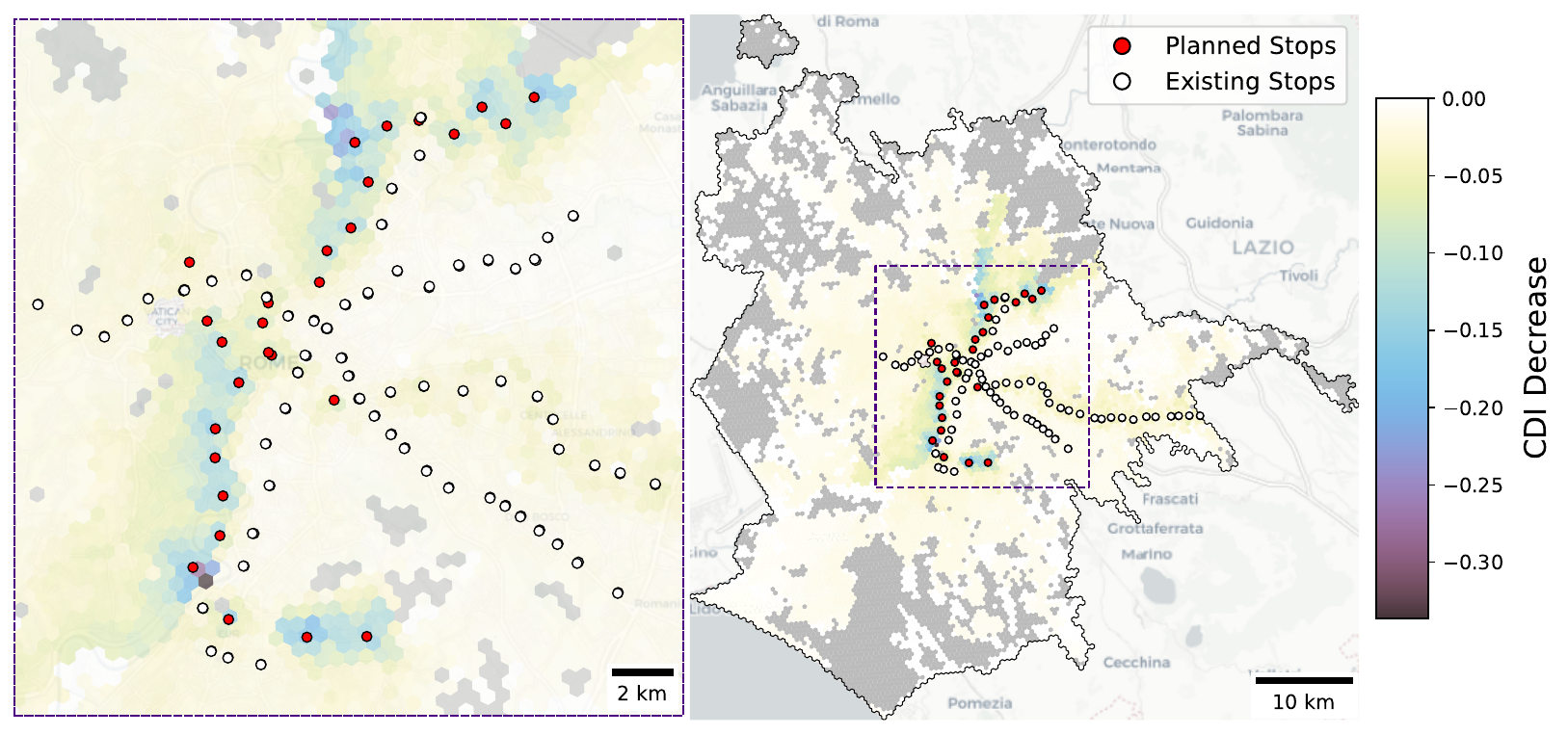}
    \caption{CDI variation in Rome after the introduction of planned new metro stops.}
    \label{fig:Rome_delta_CDI}
\end{figure*}

In Fig.~\ref{fig:Rome_delta_CDI}, we forecast the impact on car dependency of new infrastructures. Specifically, we focus here on the planned new metro stops of the C line in Rome and the opening of a planned new D line. Since the project has undergone multiple changes and has not yet been finalised, we used the last available locations at the time of writing. Notice that Rome has the largest average CDI of all cities in our sample. Naturally, the largest decrease in car dependency occurs along the new routes being built; however, there is a mild, diffuse effect across a large part of the city. Worth noting is that an appreciable decrease in CDI occurs along the \emph{old} stops of the C line, connecting notoriously underdeveloped areas in the east of the city to the centre. The city-wide variations in average CDI and opportunity score are -0.029 and 81, respectively, which become -0.058 and 187 if we consider only citizens living within a 5-minute walk (at 5 km/h) of one of the new metro stops. Therefore, at the city scale, the change would be quantitatively modest, with the entire city surpassing only Málaga in the ranking in Table~\ref{table:CDI_values}. On the other hand, if we consider the ten hexagons with the largest decrease in CDI (which are home to over 11000 citizens), the variation would be rather significant, bringing the average CDI from 0.333 to 0.110, similar to the average values for Munich or Valencia.
We then performed a logistic regression on the relationship between median CDI and car share, shown in Fig.~\ref{fig:car_use_vs_dependency}, and used the resulting logit function as a simple approximation (details in the supplementary material). We estimate that the CDI reduction could lead to an approximate 5 percentage-point decrease in the median car commuting share, from 66\% to $\sim$61\%. Since the number of employed people in Rome is approximately 1.25 million, this corresponds to over 60,000 fewer cars typically used for commuting, assuming one car per car-commuting worker. This result, although already substantial, reflects the estimated impact of a single metro line. Accordingly, the magnitude of the observed reduction highlights that achieving a meaningful, city-wide decrease in car dependency would likely require a coordinated expansion of the metro network, rather than isolated interventions.

\section*{Discussion}

In this study, we have developed a novel territorial metric to assess access to opportunities and car dependency in urban areas. By comparing the opportunities reachable by car and by public transportation, we tackled the uncomfortable question of car dependency and explored the consequences of removing cars from cities. Our analysis underscores the spatial inequalities in car dependency, offering insights into the social implications and informing more equitable urban mobility strategies. By measuring and comparing accessibility times to a range of services and facilities — denoted here generically as opportunities — for both car and public transport users, we provide a clear picture of private versus public transport in selected urban areas.

This index can be easily adapted for policymakers' use and used to evaluate the feasibility of implementing strategies to reduce car usage, such as Low Traffic Neighbourhoods and restricted traffic zones, by identifying areas where car removal is viable without significantly reducing residents' and commuters' mobility. Furthermore, the index highlights regions where public transit infrastructure needs strengthening or where establishing interchange parking areas would be beneficial.

Importantly, we show that our index correlates with car usage levels in cities, supporting the value of this methodology as a reliable indicator of car dependency. This correlation underscores the robustness of our index and highlights its applicability across different urban contexts.

In a case study on the districts of Vienna, we also measure the effect of car dependency on car ownership. To do so, we account for the effect of income, a well-established determinant of car ownership~\cite{dargay2001effect, dargay1999estimation, golob1989causal}. 

Policies also influence car ownership; for example, low-emission zones and on-street parking regulations~\cite{gonzalez2021impact}. Nevertheless, characteristics of the built environment also play an important role in shaping car ownership. It is well established that households in high-density areas tend to own fewer cars than those in low-density areas~\cite{ding2016spatial, van2010car}. One contributing factor is the greater availability of parking in low-density areas~\cite{christiansen2017household, guo2013does, Millard2020}.

However, beyond parking availability, additional mechanisms contribute to this pattern, which are captured by our car dependency index. Greater proximity to amenities—more easily achieved in denser areas—is associated with lower levels of car ownership~\cite{li2017exploring}. Moreover, high-density neighbourhoods are typically better served by public transport~\cite{giuliano2006car}, and improved job accessibility via public transport networks has been shown to play a significant role in reducing car ownership~\cite{cervero2002built}. If the abundance of points of interest (POIs) is considered a proxy for workplaces, our definition of opportunities can also capture this accessibility dimension.

Building on these observations, we proposed the car dependency index as a tool that captures some key characteristics of the urban environment associated with car ownership. Testing across the districts of Vienna, we observe a clear pattern: districts with similar mean income have very different motorisation rates, linked to residents' differing need to own a car to access opportunities in the city, as shown by the car dependency index. In particular, wealthier neighbourhoods with lower levels of car dependency, like the 7th and 8th districts, tend to own fewer cars than poorer neighbourhoods characterised by higher car dependency, like the 10th and 11th districts. This evidence sheds light on the effectiveness of reducing car dependency—particularly through the provision of efficient public transport—in lowering car ownership in urban contexts.

The map of the car dependency index in Vienna in Fig.~\ref{fig:CDI_cities} resembles the map of urban sprawl presented in~\cite{BRENNER2024107037}. This similarity suggests the realisation of the theorised feedback loop, linking car use to the generation of urban sprawl and, in turn, associating sprawled urban forms with higher levels of car dependency~\cite{galiano2021urban}. Nevertheless, further investigations would be needed to trace a causal link between these two quantities.

Last but not least, we simulate a planned metro line in Rome and estimate its impact on car dependency. While the effect is sizeable, reducing car commuting by several percentage points and resulting in tens of thousands of cars being less used for commuting, the impact is concentrated among residents living near the new stations. This spatially localised effect suggests that more systemic, network-level interventions are required to induce broader and more persistent changes in persistent mobility practices.

It is also important to note some limitations in our approach. We have simplified certain factors that influence car dependency, for instance, by using a mean-field approximation for traffic and parking times. For specific use cases, the index can be fine-tuned with refined traffic and mobility data to provide more accurate, context-specific recommendations. Despite these limitations, our findings offer a foundational framework for developing sustainable and inclusive urban mobility solutions.

\section*{Conclusions}

This study establishes the Car Dependency Index (CDI) as a robust, high-resolution metric for evaluating the structural necessity of private vehicle use in urban environments. By mathematically formalising the accessibility gap between private and public transit, our findings reveal that car dependency is not merely a behavioural choice but a spatial consequence of urban form. While central urban cores demonstrate the capacity for car-free living, peripheral districts remain trapped in high-dependency cycles, a disparity mitigated only by the presence of high-capacity rail infrastructure.

The CDI’s strong correlation with real-world commuting data and car ownership, independent of income, validates its utility as a diagnostic tool for urban planners. Our simulation of the Rome metro expansion underscores a critical policy lesson: while localised infrastructure projects yield significant benefits for immediate residents, they are insufficient to trigger city-wide shifts. To achieve durable reductions in urban emissions and address spatial inequalities, policymakers must move beyond isolated interventions toward systemic, network-level transit integration. Ultimately, the CDI provides a foundational framework for identifying where car-reduction strategies are currently viable and where massive infrastructure investment is a prerequisite for social equity.

\section*{Acknowledgments}
Hygor P. M. Melo acknowledges the support of Fundação Edson Queiroz, Universidade de Fortaleza, and Fundação Cearense de Apoio ao Desenvolvimento Científico e Tecnológico.\\ Basemaps were provided by CARTO using data from OpenStreetMap contributors. https://www.openstreetmap.org/copyright.

\section*{Data availability}

The real-world data supporting the findings of this study are publicly available from the sources cited in the manuscript. The synthetic datasets generated and analysed in this study are available from the authors upon request.

\section*{Author Contributions}
B.C.: conceptualisation, methodology, software, formal analysis, data curation, writing - original draft, writing - review and editing, visualisation.

\noindent F.M.: conceptualisation, methodology, formal analysis, writing - original draft, writing - review and editing, visualisation.

\noindent M.B.: conceptualisation, methodology, software, writing - original draft, writing - review and editing, supervision.

\noindent H.P.M.M.: conceptualisation,  writing - review and editing, visualisation, supervision.

\noindent V.L.: conceptualisation, writing - review and editing, supervision, funding acquisition.

All authors have read and approved the manuscript.

\section*{Competing Interests}
The authors declare no competing interests.

\printbibliography

@article{basilone2025,
doi = {10.1088/2632-072X/adf683},
url = {https://doi.org/10.1088/2632-072X/adf683},
year = {2025},
month = {aug},
publisher = {IOP Publishing},
volume = {6},
number = {3},
pages = {035008},
author = {Basilone, Riccardo and Bruno, Matteo and Piaget Monteiro Melo, Hygor and Avalle, Michele and Loreto, Vittorio},
title = {Road-Width-Aware network optimisation for bike lane planning},
journal = {Journal of Physics: Complexity}
}

@software{h3geo,
  title = {H3: Hexagonal Hierarchical Geospatial Indexing System},
  url = {https://h3geo.org},
}

@misc{household_size,
    url = "https://ec.europa.eu/eurostat/databrowser/view/ilc_lvph03__custom_10430677/default/table?lang=en",
    title = "Eurostat"
}

@misc{tomtom,
    title = "TomTom Traffic Index",
    url = {www.tomtom.com/traffic-index/}
}

@misc{WORLDPOP,
    title = "WorldPop",
    url = {https://hub.worldpop.org}
}

@misc{Vienna,
    title = "City of Vienna statistics",
    url = {www.wien.gv.at/statistik/verkehr-wohnen/}%,
%    urldate = {2024-10-9}
}

@misc{OSM,
    title = "OpenStreetMap",
    url = {https://www.openstreetmap.org},
}

@misc{OSRM,
    title = "Project OSRM",
    url = {https://project-osrm.org}
}

@misc{
    geofabrik,
    title = "geofabrik",
    url = {https://www.geofabrik.de}
}

@misc{
    Vienna_censo,
    title = "City of Vienna census",
    url = "https://www.wien.gv.at/statistik/bevoelkerung/"
}

@article{FUA_OECD,
   author = "Lewis Dijkstra and Hugo Poelman and Paolo Veneri",
   title = "The EU-OECD definition of a functional urban area",
   year = "2019",
   url = "https://www.oecd-ilibrary.org/content/paper/d58cb34d-en",
   doi = "https://doi.org/https://doi.org/10.1787/d58cb34d-en"
}

@article{CSA,
    title = "Connection Scan Algorithm",
    journal = "ACM Journal of Experimental Algorithmics",
    doi = "10.1145/3274661",
    author = "Julian Dibbelt, Thomas Pajor, Ben Strasser and Dorothea Wagner",
    volume = {23},
    month = oct,
    year = {2018},
    pages = {1-56},
    number = {1.7}
}

@article{dargay2001effect,
  title={The effect of income on car ownership: evidence of asymmetry},
  author={Dargay, Joyce M},
  journal={Transportation Research Part A: Policy and Practice},
  volume={35},
  number={9},
  pages={807--821},
  year={2001},
  publisher={Elsevier}
}

@article{dargay1999estimation,
  title={Estimation of a dynamic car ownership model: a pseudo-panel approach},
  author={Dargay, Joyce M and Vythoulkas, Petros C},
  journal={Journal of transport economics and policy},
  pages={287--301},
  year={1999},
  publisher={JSTOR}
}

@article{golob1989causal,
  title={The causal influences of income and car ownership on trip generation by mode},
  author={Golob, Thomas F},
  journal={Journal of Transport Economics and Policy},
  pages={141--162},
  year={1989},
  publisher={JSTOR}
}

@article{ding2016spatial,
  title={Spatial heterogeneous impact of built environment on household auto ownership levels: Evidence from analysis at traffic analysis zone scales},
  author={Ding, C and Wang, Y and Yang, J and Liu, C and Lin, Y},
  journal={Transportation Letters},
  volume={8},
  number={1},
  pages={26--34},
  year={2016},
  publisher={Taylor \& Francis}
}

@article{van2010car,
  title={Car ownership as a mediating variable in car travel behaviour research using a structural equation modelling approach to identify its dual relationship},
  author={Van Acker, Veronique and Witlox, Frank},
  journal={Journal of Transport Geography},
  volume={18},
  number={1},
  pages={65--74},
  year={2010},
  publisher={Elsevier}
}

@article{christiansen2017household,
  title={Household parking facilities: relationship to travel behaviour and car ownership},
  author={Christiansen, Petter and Fearnley, Nils and Hanssen, Jan Usterud and Skollerud, K{\aa}re},
  journal={Transportation research procedia},
  volume={25},
  pages={4185--4195},
  year={2017},
  publisher={Elsevier}
}

@article{guo2013does,
  title={Does residential parking supply affect household car ownership? The case of New York City},
  author={Guo, Zhan},
  journal={Journal of Transport Geography},
  volume={26},
  pages={18--28},
  year={2013},
  publisher={Elsevier}
}

@article{Millard2020,
    title={How the Built Environment Affects Car Ownership and Travel: Evidence from San Francisco Housing Lotteries},
    author={Millard-Ball, A. and West, J. and Rezaei, N. and Desai, G.},
    year = {2020},
    journal = {UC Office of the President: University of California Institute of Transportation Studies},
    doi = {http://dx.doi.org/10.7922/G2319T55}
}

@article{giuliano2006car,
  title={Car ownership, travel and land use: a comparison of the US and Great Britain},
  author={Giuliano, Genevieve and Dargay, Joyce},
  journal={Transportation Research Part A: Policy and Practice},
  volume={40},
  number={2},
  pages={106--124},
  year={2006},
  publisher={Elsevier}
}

@article{li2017exploring,
  title={Exploring car ownership and car use in neighborhoods near metro stations in Beijing: Does the neighborhood built environment matter?},
  author={Li, Shengxiao and Zhao, Pengjun},
  journal={Transportation research part D: transport and environment},
  volume={56},
  pages={1--17},
  year={2017},
  publisher={Elsevier}
}

@article{cervero2002built,
  title={Built environments and mode choice: toward a normative framework},
  author={Cervero, Robert},
  journal={Transportation Research Part D: Transport and Environment},
  volume={7},
  number={4},
  pages={265--284},
  year={2002},
  publisher={Elsevier}
}

@article{blumenberg2014driving,
  title={A driving factor in mobility? Transportation's role in connecting subsidized housing and employment outcomes in the moving to opportunity (MTO) program},
  author={Blumenberg, Evelyn and Pierce, Gregory},
  journal={Journal of the American Planning Association},
  volume={80},
  number={1},
  pages={52--66},
  year={2014},
  publisher={Taylor \& Francis}
}

@article{rueda2019superblocks,
  title={Superblocks for the design of new cities and renovation of existing ones: Barcelona’s case},
  author={Rueda, Salvador},
  journal={Integrating human health into urban and transport planning: A framework},
  pages={135--153},
  year={2019},
  publisher={Springer}
}

@article{prud2005london,
  title={The London congestion charge: a tentative economic appraisal},
  author={Prud'Homme, Remy and Bocarejo, Juan Pablo},
  journal={Transport Policy},
  volume={12},
  number={3},
  pages={279--287},
  year={2005},
  publisher={Elsevier}
}

@article{borjesson2012stockholm,
  title={The Stockholm congestion charges—5 years on. Effects, acceptability and lessons learnt},
  author={B{\"o}rjesson, Maria and Eliasson, Jonas and Hugosson, Muriel B and Brundell-Freij, Karin},
  journal={Transport Policy},
  volume={20},
  pages={1--12},
  year={2012},
  publisher={Elsevier}
}

@article{ostermeijer2019residential,
  title={Residential parking costs and car ownership: Implications for parking policy and automated vehicles},
  author={Ostermeijer, Francis and Koster, Hans RA and van Ommeren, Jos},
  journal={Regional Science and Urban Economics},
  volume={77},
  pages={276--288},
  year={2019},
  publisher={Elsevier}
}

@article{gossling2020cities,
  title={Why cities need to take road space from cars-and how this could be done},
  author={G{\"o}ssling, Stefan},
  journal={Journal of Urban Design},
  volume={25},
  number={4},
  pages={443--448},
  year={2020},
  publisher={Taylor \& Francis}
}

@article{nieuwenhuijsen2016car,
  title={Car free cities: Pathway to healthy urban living},
  author={Nieuwenhuijsen, Mark J and Khreis, Haneen},
  journal={Environment international},
  volume={94},
  pages={251--262},
  year={2016},
  publisher={Elsevier}
}

@article{gonzalez2021impact,
  title={What impact do private vehicle restrictions in urban areas have on car ownership? Empirical evidence from the city of Madrid},
  author={Gonzalez, Juan Nicolas and Perez-Doval, Jose and Gomez, Juan and Vassallo, Jose Manuel},
  journal={Cities},
  volume={116},
  pages={103301},
  year={2021},
  publisher={Elsevier}
}

@article{PRIETOCURIEL2024ABC,
title = {The ABC of mobility},
journal = {Environment International},
volume = {185},
pages = {108541},
year = {2024},
issn = {0160-4120},
doi = {https://doi.org/10.1016/j.envint.2024.108541},
author = {Rafael Prieto-Curiel and Juan P. Ospina}
}

@techreport{IEA2023,

title={Net Zero
Roadmap.
A Global Pathway to
Keep the 1.5 °C Goal
in Reach - 2023 Update}, 
institution = {IEA},
url = {www.iea.org/reports/net-zero-roadmap-a-global-pathway-to-keep-the-15-0c-goal-in-reach},
urldate = {2024-10-03}
}

@techreport{NetZero2050,
    title = {Net {{Zero}} by 2050 - {{A Roadmap}} for the {{Global Energy Sector}}},
    institution = {IEA},
    url={www.iea.org/reports/net-zero-by-2050},
    urldate = {2024-10-03}
}

@article{wallington2022vehicle,
  title={Vehicle emissions and urban air quality: 60 years of progress},
  author={Wallington, Timothy J and Anderson, James E and Dolan, Rachael H and Winkler, Sandra L},
  journal={Atmosphere},
  volume={13},
  number={5},
  pages={650},
  year={2022},
  publisher={MDPI}
}

@book{verkade2024movement,
  title={Movement: how to take back our streets and transform our lives},
  author={Verkade, Thalia and Te Br{\"o}mmelstroet, Marco},
  year={2024},
  publisher={Island Press}
}

@book{peden2004world,
  title={World report on road traffic injury prevention},
  author={Peden, Margie M},
  year={2004},
  publisher={World Health Organization}
}

@article{galiano2021urban,
  title={Urban sprawl and mobility},
  author={Galiano, Giuseppe and Forestieri, Giulia and Moretti, Laura and others},
  journal={WIT Trans. Built Environ},
  volume={204},
  pages={245--255},
  year={2021}
}

@article{BRENNER2024107037,
title = {What drives densification and sprawl in cities? A spatially explicit assessment for Vienna, between 1984 and 2018},
journal = {Land Use Policy},
volume = {138},
pages = {107037},
year = {2024},
issn = {0264-8377},
doi = {https://doi.org/10.1016/j.landusepol.2023.107037},
url = {https://www.sciencedirect.com/science/article/pii/S0264837723005033},
author = {Anna-Katharina Brenner and Willi Haas and Tobias Krüger and Sarah Matej and Helmut Haberl and Franz Schug and Dominik Wiedenhofer and Martin Behnisch and Jochen A.G. Jaeger and Melanie Pichler}
}

@article{levinsonGeneralTheoryAccess2020,
  title = {Towards a General Theory of Access},
  author = {Levinson, David M and Wu, Hao},
  year = {2020},
  month = jun,
  journal = {Journal of Transport and Land Use},
  volume = {13},
  number = {1},
  pages = {129--158},
  issn = {1938-7849},
  doi = {10.5198/jtlu.2020.1660},
  urldate = {2024-03-27},
  copyright = {https://creativecommons.org/licenses/by-nc-nd/4.0},
  langid = {english}
}

@article{loufHowCongestionShapes2014,
  title = {How Congestion Shapes Cities: From Mobility Patterns to Scaling},
  shorttitle = {How Congestion Shapes Cities},
  author = {Louf, R{\'e}mi and Barthelemy, Marc},
  year = {2014},
  month = jul,
  journal = {Scientific Reports},
  volume = {4},
  number = {1},
  pages = {5561},
  issn = {2045-2322},
  doi = {10.1038/srep05561},
  urldate = {2023-11-21},
  langid = {english}
}

@article{brommelstroetIdentifyingNurturingEmpowering2022,
  title = {Identifying, Nurturing and Empowering Alternative Mobility Narratives},
  author = {Br{\"o}mmelstroet, Marco Te and Mladenovi{\'c}, Milo{\v s} N. and Nikolaeva, Anna and Gaziulusoy, {\.I}dil and Ferreira, Antonio and {Schmidt-Thom{\'e}}, Kaisa and Ritvos, Roope and Sousa, Silvia and Bergsma, Bernadette},
  year = {2022},
  month = dec,
  journal = {Journal of Urban Mobility},
  volume = {2},
  pages = {100031},
  issn = {26670917},
  doi = {10.1016/j.urbmob.2022.100031}
}

@article{bruno2024universal,
  title={A universal framework for inclusive 15-minute cities},
  author={Bruno, Matteo and Monteiro Melo, Hygor Piaget and Campanelli, Bruno and Loreto, Vittorio},
  journal={Nature Cities},
  volume={1},
  number={10},
  pages={633--641},
  year={2024},
  publisher={Nature Publishing Group US New York}
}

@article{rhoadsInclusive15minuteCity2023,
  title = {The Inclusive 15-Minute City: {{Walkability}} Analysis with Sidewalk Networks},
  shorttitle = {The Inclusive 15-Minute City},
  author = {Rhoads, Daniel and {Sol{\'e}-Ribalta}, Albert and {Borge-Holthoefer}, Javier},
  year = {2023},
  month = mar,
  journal = {Computers, Environment and Urban Systems},
  volume = {100},
  pages = {101936},
  issn = {01989715},
  doi = {10.1016/j.compenvurbsys.2022.101936},
  urldate = {2024-09-04},
  langid = {english}
}

@article{hill2024beyond,
  author = {Dan Hill and Matteo Bruno and Hygor P. M. Melo and Yuichiro Takeuchi and Vittorio Loreto},
  title = {Cities beyond proximity},
  journal = {Philosophical Transactions of the Royal Society A},
  volume = {382},
  number = {20240097},
  year = {2024},
  url = {http://doi.org/10.1098/rsta.2024.0097},
  doi = {10.1098/rsta.2024.0097}
}

@article{abbiasov202415,
  title={The 15-minute city quantified using human mobility data},
  author={Abbiasov, Timur and Heine, Cate and Sabouri, Sadegh and Salazar-Miranda, Arianna and Santi, Paolo and Glaeser, Edward and Ratti, Carlo},
  journal={Nature Human Behaviour},
  volume={8},
  number={3},
  pages={445--455},
  year={2024},
  publisher={Nature Publishing Group UK London}
}

@article{marzolla2024compact,
  title={Compact 15-minute cities are greener},
  author={Marzolla, Francesco and Bruno, Matteo and Melo, Hygor Piaget Monteiro and Loreto, Vittorio},
  journal={arXiv preprint arXiv:2409.01817},
  year={2024}
}

@article{hansen1959accessibility,
  title={How accessibility shapes land use},
  author={Hansen, Walter G},
  journal={Journal of the American Institute of planners},
  volume={25},
  number={2},
  pages={73--76},
  year={1959},
  publisher={Taylor \& Francis}
}

@article{wu2021urban,
  title={Urban access across the globe: an international comparison of different transport modes},
  author={Wu, Hao and Avner, Paolo and Boisjoly, Genevieve and Braga, Carlos KV and El-Geneidy, Ahmed and Huang, Jie and Kerzhner, Tamara and Murphy, Brendan and Niedzielski, Micha{\l} A and Pereira, Rafael HM and others},
  journal={npj Urban Sustainability},
  volume={1},
  number={1},
  pages={16},
  year={2021},
  publisher={Nature Publishing Group UK London}
}

@article{albalate2020impact,
  title={The impact of curbside parking regulations on car ownership},
  author={Albalate, Daniel and Gragera, Albert},
  journal={Regional Science and Urban Economics},
  volume={81},
  pages={103518},
  year={2020},
  publisher={Elsevier}
}

@article{glazener2022impacts,
  title={The impacts of car-free days and events on the environment and human health},
  author={Glazener, Andrew and Wylie, James and van Waas, Willem and Khreis, Haneen},
  journal={Current Environmental Health Reports},
  volume={9},
  number={2},
  pages={165--182},
  year={2022},
  publisher={Springer}
}

@article{folco2023data,
  title={Data-driven micromobility network planning for demand and safety},
  author={Folco, Pietro and Gauvin, Laetitia and Tizzoni, Michele and Szell, Michael},
  journal={Environment and planning B: Urban analytics and city science},
  volume={50},
  number={8},
  pages={2087--2102},
  year={2023},
  publisher={SAGE Publications Sage UK: London, England}
}

@article{metz2008myth,
  title={The myth of travel time saving},
  author={Metz, David},
  journal={Transport reviews},
  volume={28},
  number={3},
  pages={321--336},
  year={2008},
  publisher={Taylor \& Francis}
}

@book{newman1989cities,
  title={Cities and automobile dependence: An international sourcebook},
  author={Newman, Peter G and Kenworthy, Jeffrey R},
  year={1989}
}

@article{lucas2012transport,
  title={Transport and social exclusion: Where are we now?},
  author={Lucas, Karen},
  journal={Transport policy},
  volume={20},
  pages={105--113},
  year={2012},
  publisher={Elsevier}
}

@article{urry2004system,
  title={The ‘system’of automobility},
  author={Urry, John},
  journal={Theory, culture \& society},
  volume={21},
  number={4-5},
  pages={25--39},
  year={2004},
  publisher={Sage London, Thousand Oaks and New Delhi}
}

@article{wiersma2021spatial,
  title={Spatial conditions for car dependency in mid-sized European city regions},
  author={Wiersma, JK and Bertolini, L and Harms, L},
  journal={European Planning Studies},
  volume={29},
  number={7},
  pages={1314--1330},
  year={2021},
  publisher={Taylor \& Francis}
}

@article{chang2017there,
  title={Is there more traffic congestion in larger cities?-Scaling analysis of the 101 largest US urban centers},
  author={Chang, Yu Sang and Lee, Yong Joo and Choi, Sung Sup Brian},
  journal={Transport Policy},
  volume={59},
  pages={54--63},
  year={2017},
  publisher={Elsevier}
}

@article{han2022effect,
  title={The effect of commuting time on quality of life: Evidence from China},
  author={Han, Libin and Peng, Chong and Xu, Zhenyu},
  journal={International journal of environmental research and public health},
  volume={20},
  number={1},
  pages={573},
  year={2022},
  publisher={MDPI}
}

@article{bruno2026dimensions,
  author    = {Matteo Bruno and Bruno Campanelli and Hygor P. M. Melo and Lavinia Rossi Mori and Vittorio Loreto},
  title     = {The dimensions of accessibility: proximity, opportunities, values},
  journal   = {EPJ Data Science},
  year      = {2026},
  month     = {jan},
  day       = {28},
  issn      = {2193-1127},
  doi       = {10.1140/epjds/s13688-026-00623-8},
  url       = {https://doi.org/10.1140/epjds/s13688-026-00623-8}
}

@article{bertolini1999spatial,
  title={Spatial development patterns and public transport: the application of an analytical model in the Netherlands},
  author={Bertolini, Luca},
  journal={Planning practice and research},
  volume={14},
  number={2},
  pages={199--210},
  year={1999},
  publisher={Taylor \& Francis}
}

@article{marzolla2026proximity,
  title={Proximity-based cities emit less mobility-driven CO2},
  author={Marzolla, Francesco and M. Melo, Hygor P and Bruno, Matteo and Loreto, Vittorio},
  journal={npj Sustainable Mobility and Transport},
  volume={3},
  number={1},
  pages={7},
  year={2026},
  publisher={Nature Publishing Group UK London}
}

@article{mattioli2020political,
  title={The political economy of car dependence: A systems of provision approach},
  author={Mattioli, Giulio and Roberts, Cameron and Steinberger, Julia K and Brown, Andrew},
  journal={Energy research \& social science},
  volume={66},
  pages={101486},
  year={2020},
  publisher={Elsevier}
}

@report{kiberd2024CarDependency,
  title        = {Trapped Behind the Wheel: How Neighbourhoods in England Became Car Dependent and What to Do About It},
  author       = {Kiberd, Emmet and Straňák, Benedikt},
  year         = {2024},
  institution  = {New Economics Foundation},
  address      = {London, UK},
  url          = {https://neweconomics.org/2024/11/trapped-behind-the-wheel},
}

@article{mattioli2017forced,
  title={‘Forced Car Ownership’in the UK and Germany: socio-spatial patterns and potential economic stress impacts},
  author={Mattioli, Giulio},
  journal={Social Inclusion},
  volume={5},
  number={4},
  pages={147--160},
  year={2017}
}
\end{multicols}

\end{document}


\maketitle

\section*{Border Effects}

Every city in this work is defined through its administrative boundaries. However, people residing in hexagons near the boundary
should be expected to be able to access POIs that are outside the city. Not considering the latter might therefore artificially lower
the opportunity scores of peripheral areas. On the other hand, it is often not possible to reliably obtain interurban public
transport schedules. In this section we consider the example of Paris, for which schedules outside the administrative boundaries
are available, to show that the effect is small and does not qualitatively change our assessment of the city's accessibility patterns.

We recomputed the metrics for Paris by using the much larger OECD urban core city limits, removed all hexagons
that are outside the administrative boundary, and finally compared the results with the ones shown in the main paper.
For the sake of readability, we use ``open" and ``closed" to refer respectively to the two sets of results, since in the first
case it is possible to search for POIs beyond the administrative boundaries while in the second one it is not.

In table \ref{table:correlations} we show Pearson's and Spearman's correlation coefficient between the metrics for the open
and  closed city. The values in the rows labelled as ``Close to Boundary" are computed only over hexagons that are at most
350 m away from the administrative boundary. The relative ordering of CDI values changes very little. The only large change is in the opportunity score by car. This is due to the centre
of Paris being very much a pedestrian-first zone, so opening the border means allowing exploration of more car-friendly areas, and the hexagons near the border end up having a higher opportunity score by car than the ones in the centre. This effect still leaves very
high $r$ and $\rho$ for the CDI, and furthermore we don't expect to see it in other cities.

\begin{table}[H]
\centering
\setlength{\tabcolsep}{6pt} 
\renewcommand{\arraystretch}{1.2} 
\caption{\textbf{Correlations between metrics for ``open" and ``closed" Paris}}
\begin{tabular}{cccc}
\textbf{Metric} & \textbf{Hexagon Set} & \textbf{Pearson $r$} & \textbf{Spearman $\rho$}  \\
\hline
Opportunity Score (PT) & All & 0.97 & 0.98  \\
Opportunity Score (Car) & All & 0.43 & 0.23 \\
CDI & All & 0.87 & 0.94 \\
Opportunity Score (PT) & Close to Boundary & 0.92 & 0.95  \\
Opportunity Score (Car) & Close to Boundary & 0.87 & 0.91 \\
CDI & Close to Boundary & 0.82 & 0.92 \\
\end{tabular}
\label{table:correlations}
\end{table}

\section*{Accessibility metrics for all cities}

\begin{figure}[H]
\centering
    \includegraphics[height=.75\paperheight]{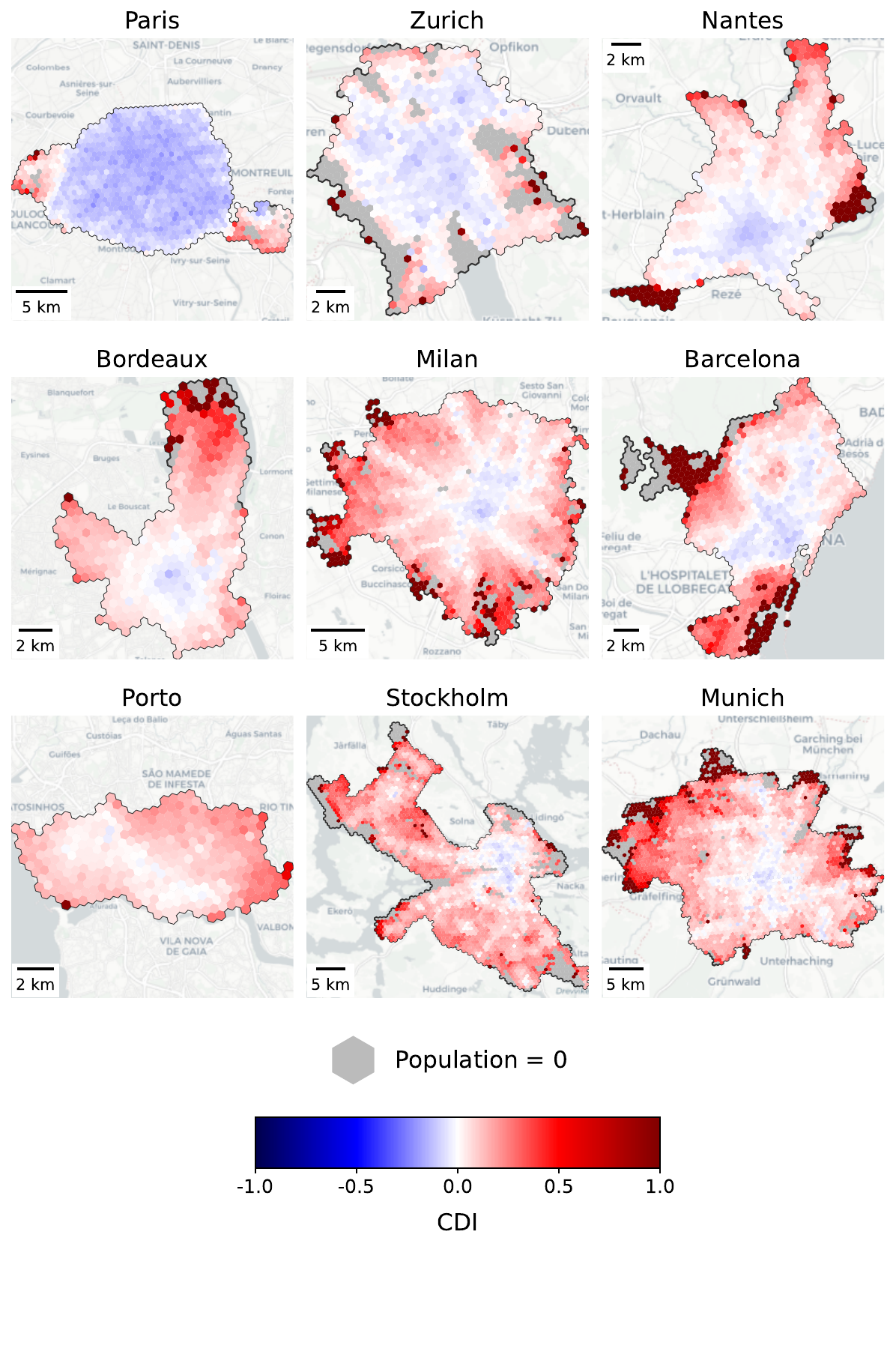}
    \caption{Car Dependency Index scores for cities under study. A positive score means that a car can access more opportunities than public transportation, vice versa for a negative score.}
\label{fig:cdi_1}
\end{figure}

\begin{figure}[H]
\centering
    \includegraphics[height=.75\paperheight]{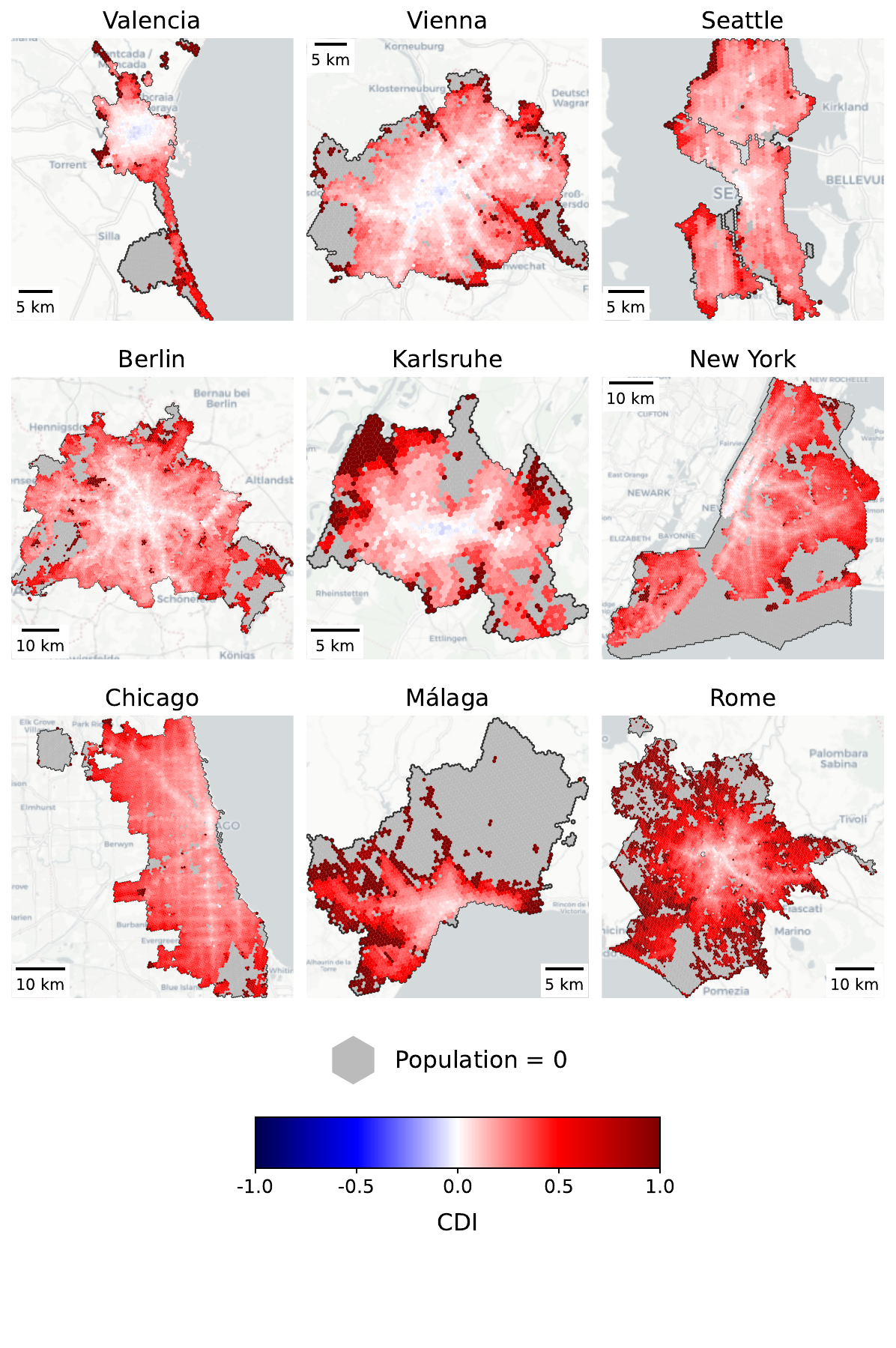}
    \caption{Car Dependency Index scores for cities under study. A positive score means that a car can access more opportunities than public transportation, vice versa for a negative score.}
\label{fig:cdi_2}
\end{figure}

\begin{figure}[H]
\centering
    \includegraphics[height=.75\paperheight]{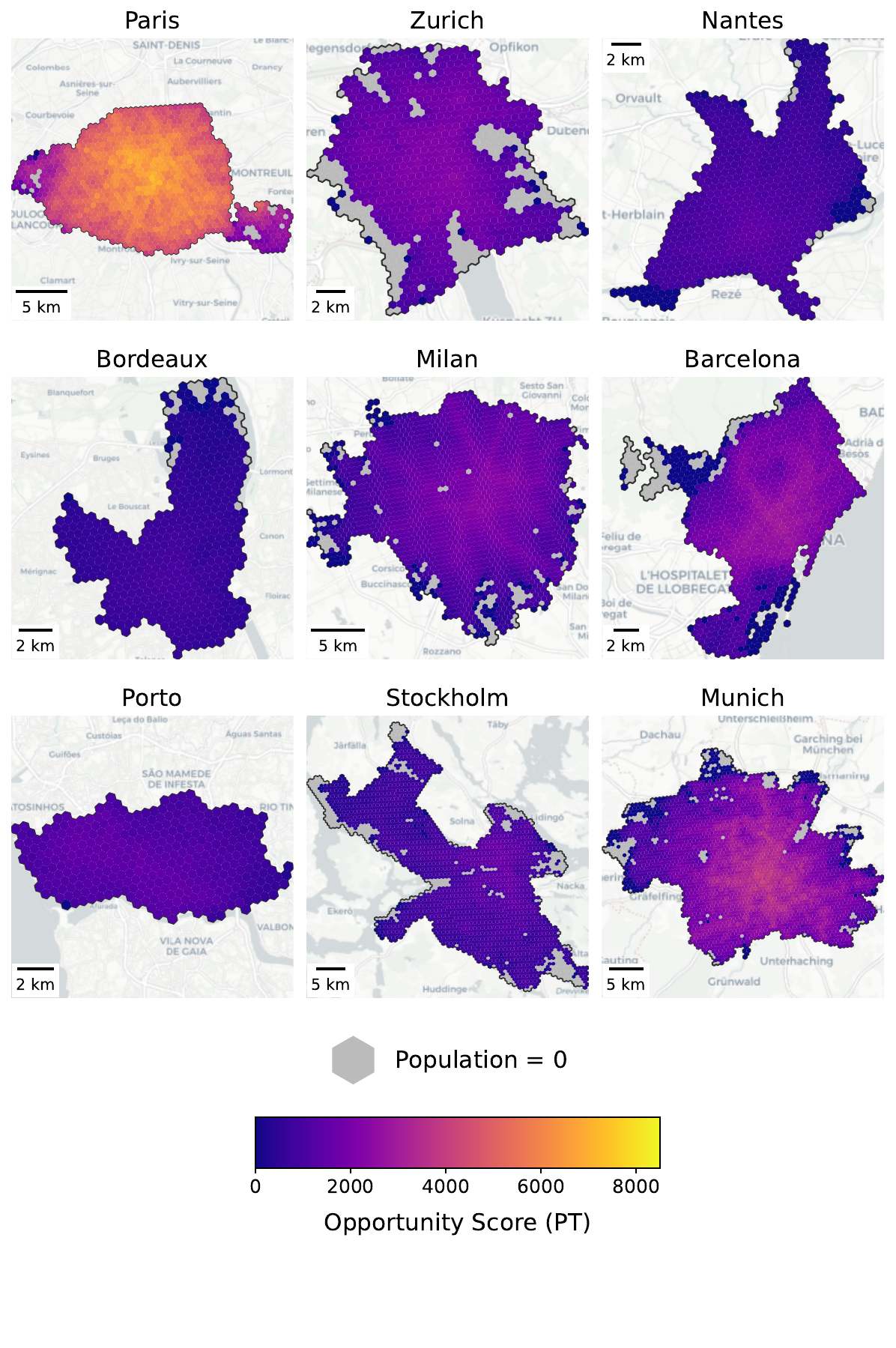}
    \caption{Opportunity scores computed for public transportation for cities under study. Opportunity scores measure the number of POIs of various types reachable with a journey of a reasonable time.}
\label{fig:opp_pt_1}
\end{figure}

\begin{figure}[H]
\centering
    \includegraphics[height=.75\paperheight]{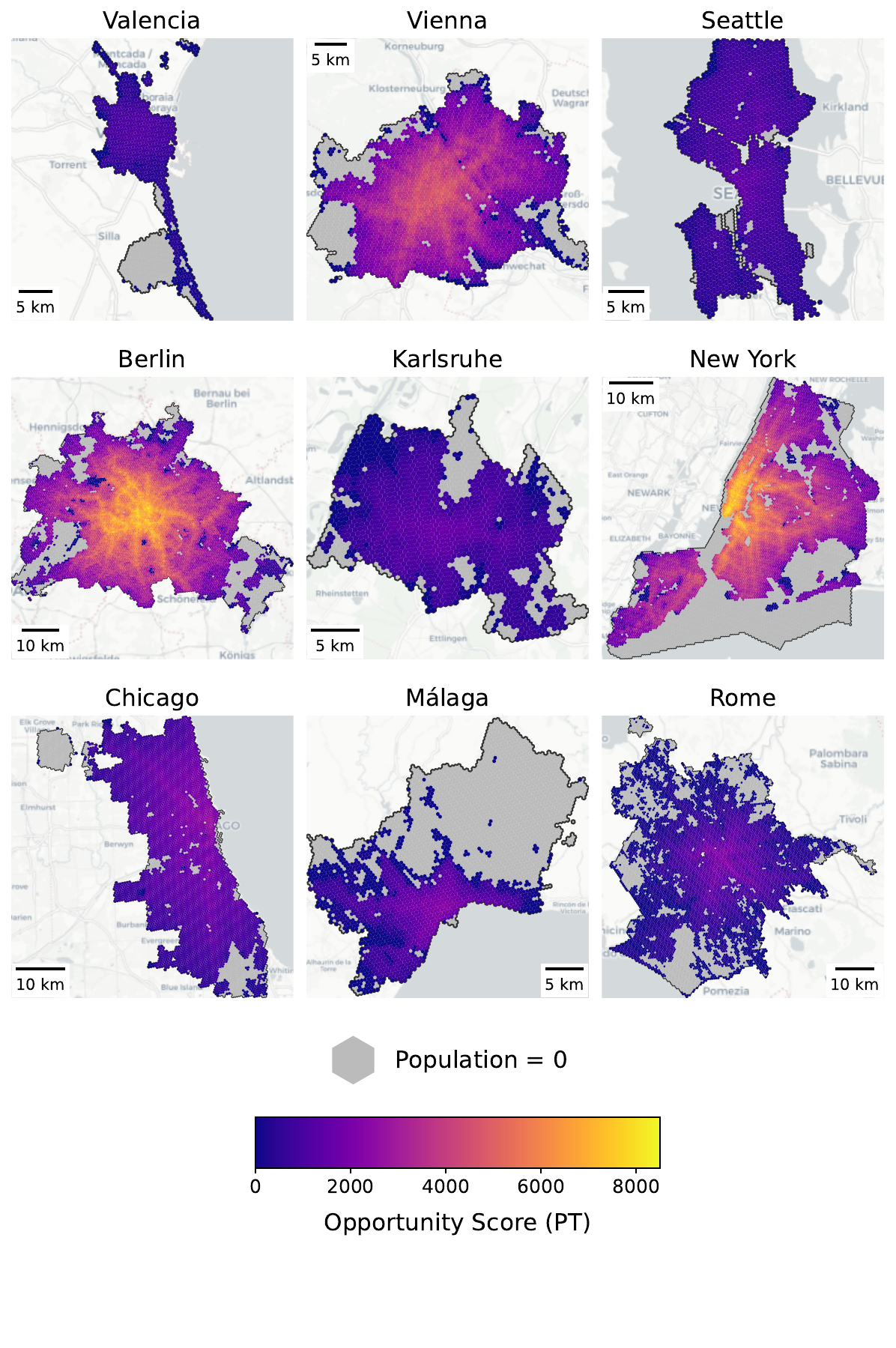}
    \caption{Opportunity scores computed for public transportation for cities under study. Opportunity scores measure the number of POIs of various types reachable with a journey of a reasonable time.}
\label{fig:opp_pt_2}
\end{figure}

\begin{figure}[H]
\centering
    \includegraphics[height=.75\paperheight]{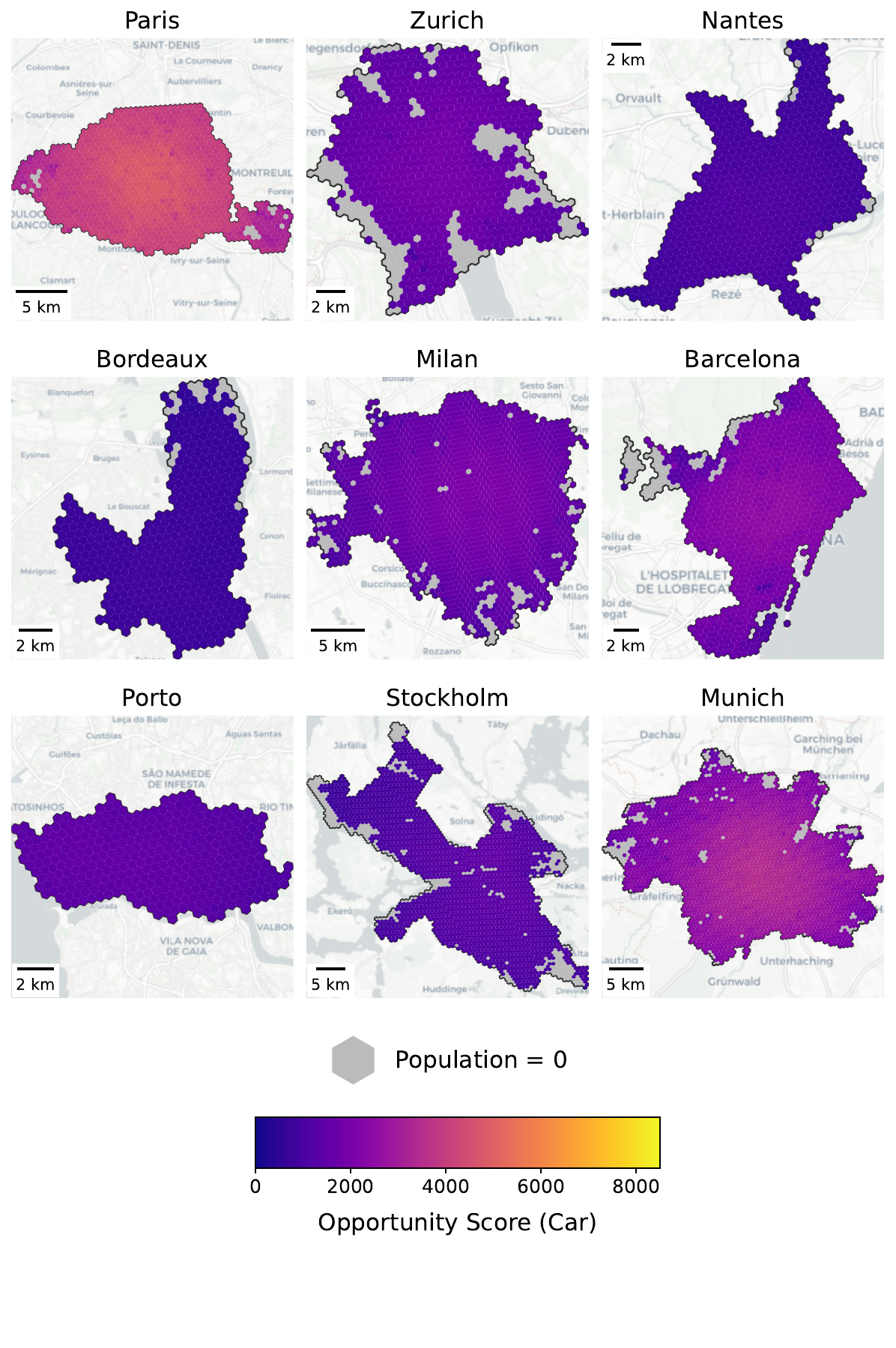}
    \caption{Opportunity scores computed for car for cities under study. Opportunity scores measure the number of POIs of various types reachable with a journey of a reasonable time.}
\label{fig:opp_car_1}
\end{figure}

\begin{figure}[H]
\centering
    \includegraphics[height=.75\paperheight]{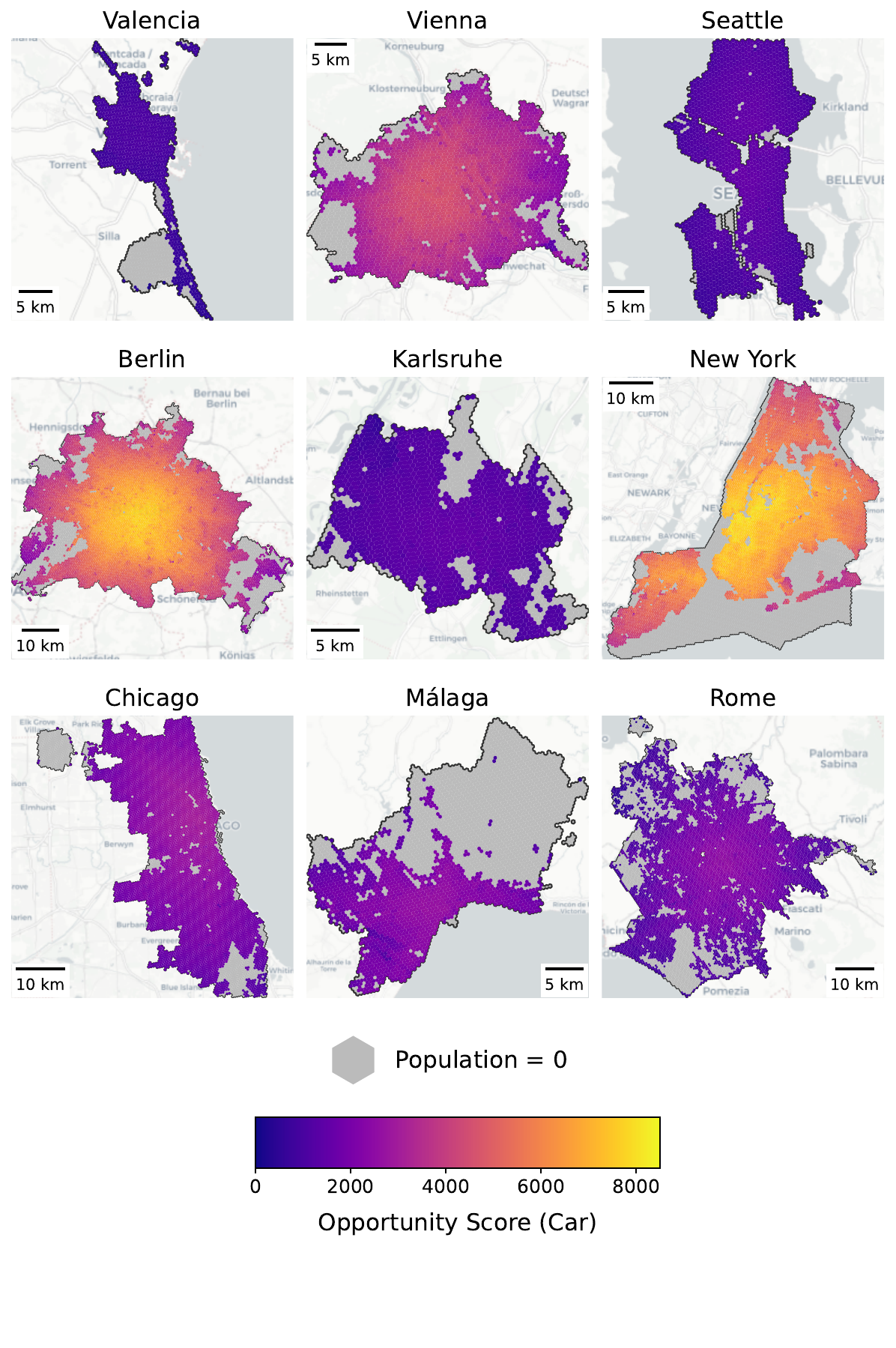}
    \caption{Opportunity scores computed for car for cities under study. Opportunity scores measure the number of POIs of various types reachable with a journey of a reasonable time.}
 \label{fig:opp_car_2}
\end{figure}

Figures \ref{fig:cdi_1}--\ref{fig:opp_car_2} show the accessibility metrics at the local level
for all cities in our samples: car dependency index (figures \ref{fig:cdi_1} and \ref{fig:cdi_2}), opportunity score by public transport (figures \ref{fig:opp_pt_1} and
\ref{fig:opp_pt_2}) and opportunity score by car (figures \ref{fig:opp_car_1} and \ref{fig:opp_car_2}). 

\section*{Comparison of different utility functions}

\begin{figure}[H] 
\centering
    \includegraphics[width=.9\linewidth]{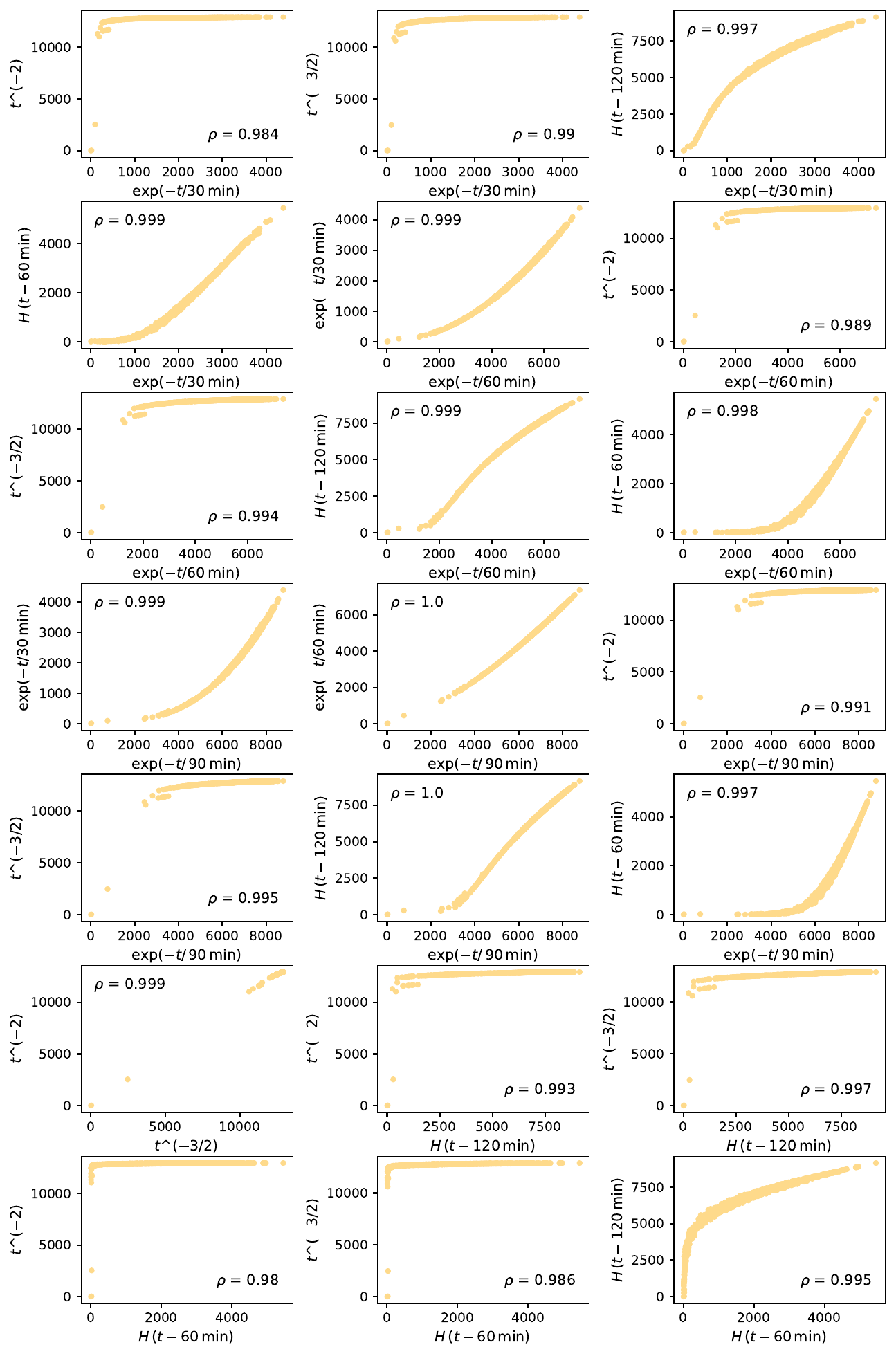}
    \caption{Relationship between opportunity score (PT) values calculated for Paris with different utility functions. Graph labels do not include normalization factors for visual clarity. $H$ is the Heaviside function, $\rho$ is Spearman's rank correlation coefficient.}
    \label{fig:util_functions_comparison_PT}
\end{figure}

\begin{figure}[H] 
\centering
    \includegraphics[width=.9\linewidth]{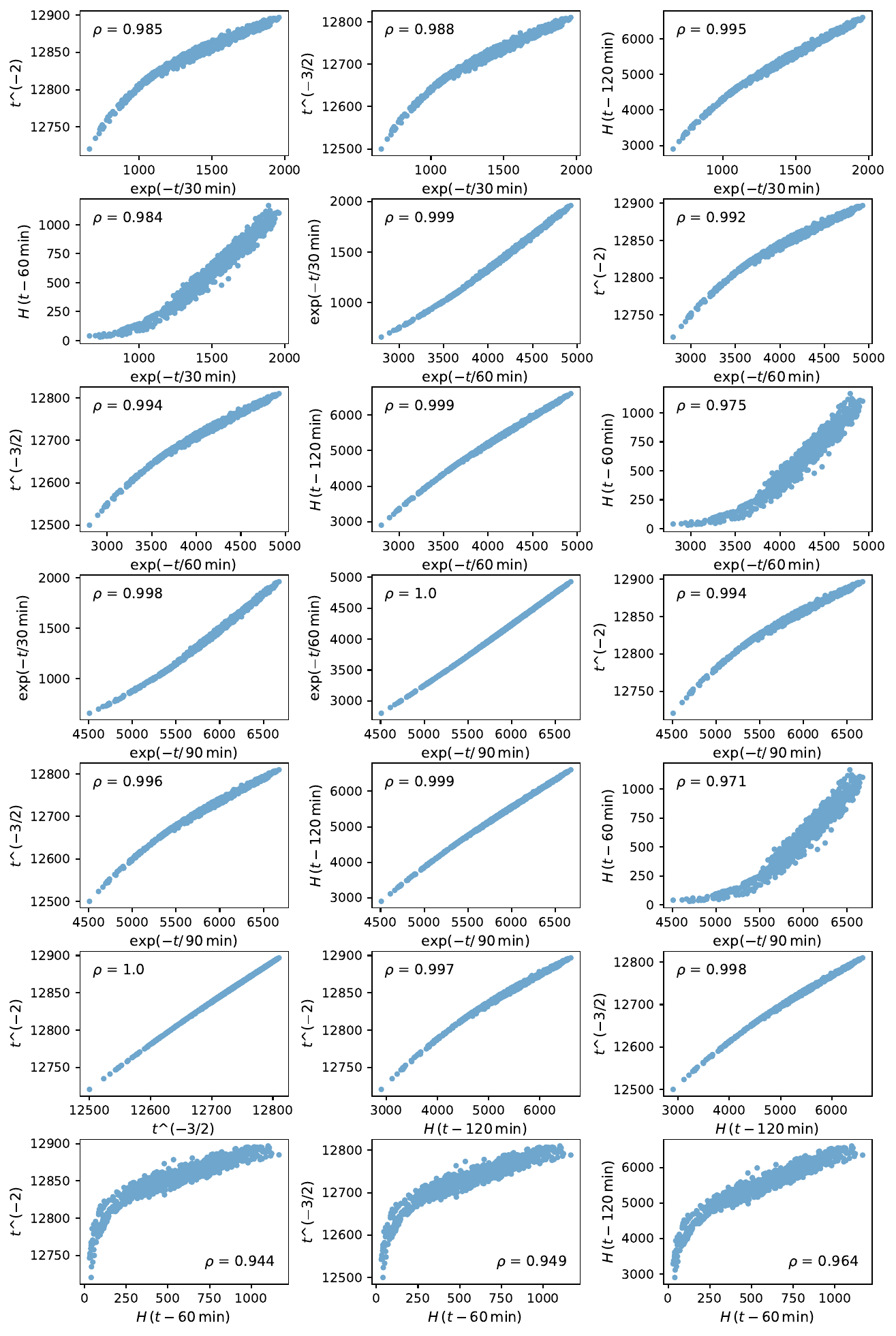}
    \caption{Relationship between opportunity score (car) values calculated for Paris with different utility functions. Graph labels do not include normalization factors for visual clarity. $H$ is the Heaviside function, $\rho$ is Spearman's rank correlation coefficient.}
    \label{fig:util_functions_comparison_car}
\end{figure}

\begin{figure}[H] 
\centering
    \includegraphics[width=.9\linewidth]{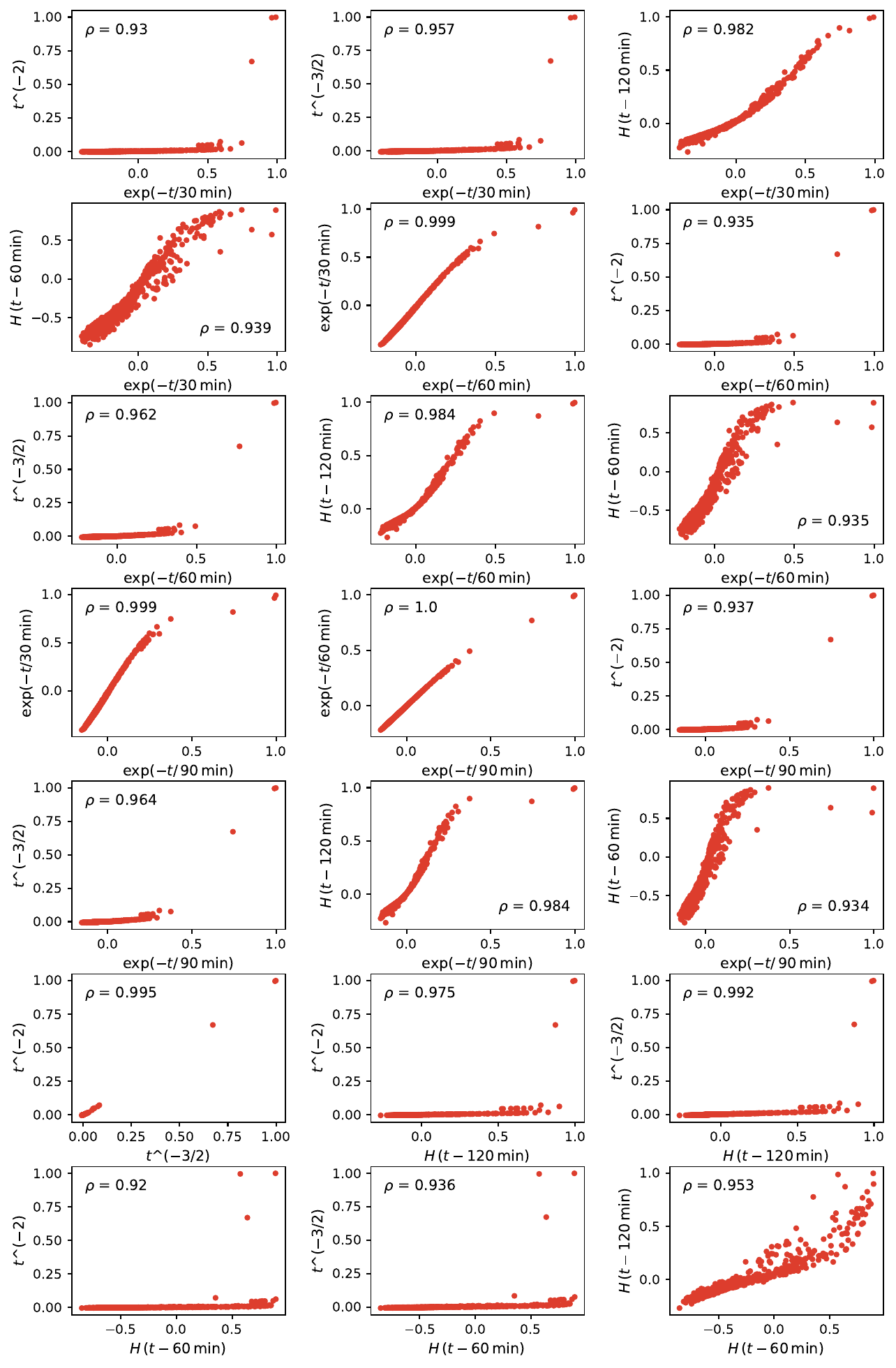}
    \caption{Relationship between CDI values calculated for Paris with different utility functions. Graph labels do not include normalization factors for visual clarity. $H$ is the Heaviside function, $\rho$ is Spearman's rank correlation coefficient.}
    \label{fig:util_functions_comparison_CDI}
\end{figure}

In Figures \ref{fig:util_functions_comparison_PT}, \ref{fig:util_functions_comparison_car} and \ref{fig:util_functions_comparison_CDI}, we check if any noticeable difference make the Car Dependency Index dependent on the chosen function by comparing different utility functions used in the computation of the opportunity scores. We test the differences on Paris using exponential, power law, and step function decays, with different parameters. For opportunity scores for public transport and car, the Spearman rank correlation between any pair of metric is never lower than 0.94. The resulting rankings of CDI scores are also very much metric-independent, as the lowest Spearman coefficient is 0.92. Some inevitable outliers are present, mostly due to car dependent places that have services at specific distances.

\section*{Comparison of different parking times}

\begin{figure}[H] 
\centering
    \includegraphics[width=.9\linewidth]{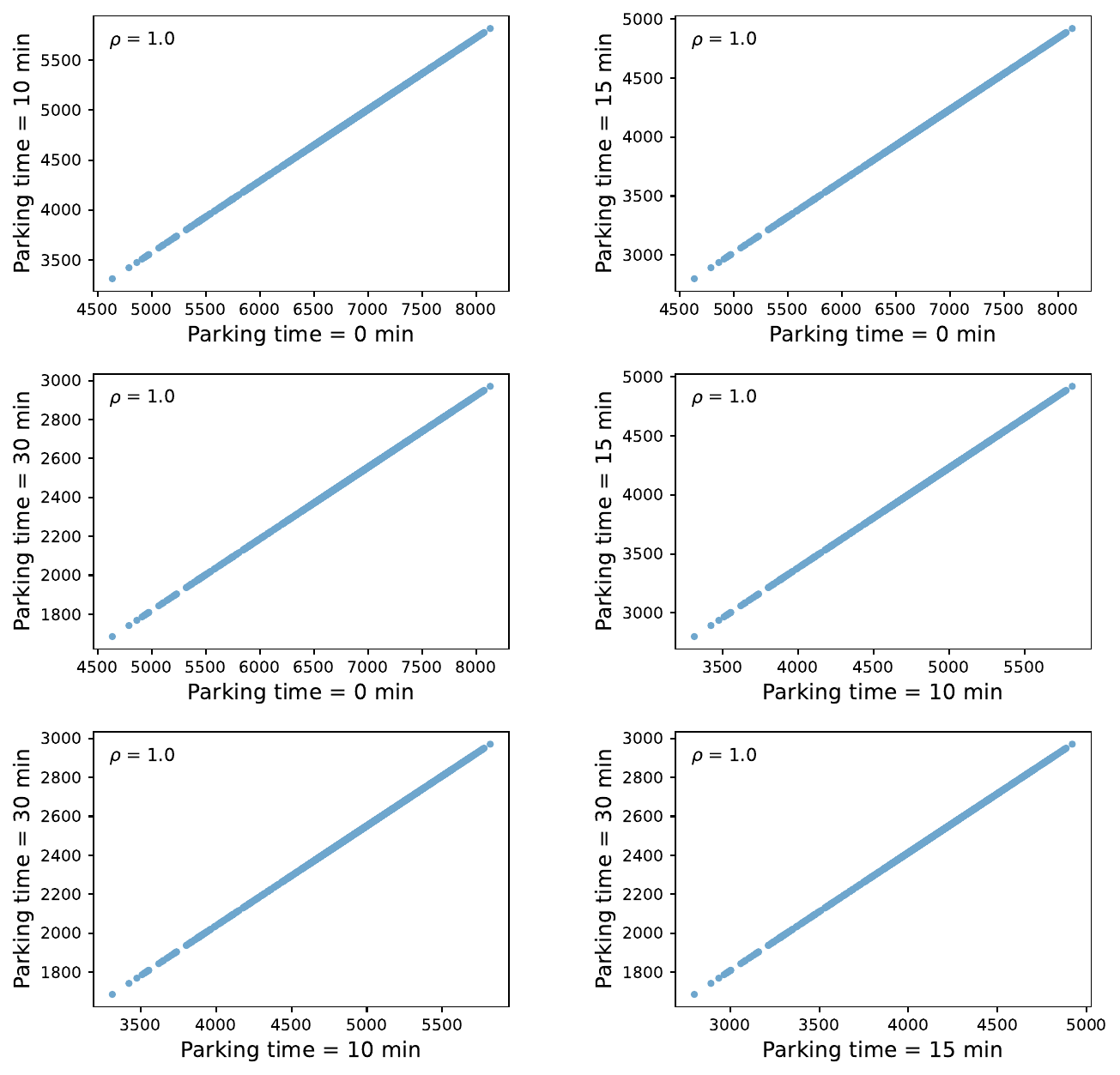}
    \caption{Relationship between opportunity scores (car) computed for Paris with different parking times. $\rho$ is Spearman's rank correlation coefficient. The relationships between CDI values are not shown since $\text{CDI}_h$ is a monotonic function of $O_{h, \text{Car}}$ for fixed $O_{h, \text{PT}}$.}
    \label{fig:parking_comparison}
\end{figure}

In Figure \ref{fig:parking_comparison}, we compare different choices of parking times for Paris. The opportunity score by car is correlated with the chosen duration of the parking time and the time is added uniformly in the city, therefore the rankings are always identical.

\pagebreak

